\pdfoutput=1
\documentclass[10pt]{article}
\usepackage{epsfig,amsmath,float,fullpage}
\usepackage{url}
\usepackage{cite}
\usepackage{sidecap}
\usepackage{helvet}
\usepackage{multirow}
\usepackage[pdftex]{color}
\usepackage{graphicx}
\usepackage{amsthm}
\usepackage{amssymb}
\usepackage{epstopdf}
\usepackage{hyperref}
\usepackage{pifont}
\usepackage[multiple,hang]{footmisc}

\newcommand{\R}{\textsf{R}}
\newcommand{\cranr}{\textsf{CRAN-R}}
\newcommand{\baysurv}{\texttt{bayessurvreg1()}}
\newcommand{\bgp}{\texttt{BeMRes()}}
\newcommand{\lddp}{\texttt{LDDPsurvival()}}
\newcommand{\emrh}{\texttt{estimateMRH()}}
\newcommand{\bshaz}{\texttt{bshazard()}}
\newcommand{\muhaz}{\texttt{muhaz()}}
\newcommand{\presmooth}{\texttt{presmooth()}}
\newcommand{\timecox}{\texttt{timecox()}}
\newcommand{\ypmodel}{\texttt{YPmodel.IntervalBands()}}
\newcommand{\makefxn}[1]{\texttt{#1()}}

\begin{document}

\title{Comparison of hazard rate estimation in \R{}}

\renewcommand{\arraystretch}{0.65}
\author{\small Yolanda Hagar and Vanja Dukic \footnote{
Yolanda Hagar  is a postdoctoral researcher and Vanja Dukic is an Associate Professor in Applied Mathematics, University of Colorado at Boulder.  Correspondence emails: {\tt yolanda.hagar@colorado.edu, vanja.dukic@colorado.edu.}
}}

\date{  }
\maketitle

\thispagestyle{empty}
\baselineskip 12pt

\begin{abstract}
\noindent
We give an overview of eight different software packages and functions available in  \R{} for semi- or non-parametric estimation of the hazard rate for right-censored survival data.  Of particular interest is the accuracy of the estimation of the hazard rate in the presence of covariates, as well as the user-friendliness of the packages.  In addition, we investigate the ability to incorporate covariates under both the proportional and the non-proportional hazards assumptions.  We contrast the robustness, variability and precision of the functions through simulations, and then further compare differences between the functions by analyzing the ``cancer" and ``TRACE" survival data sets available in  \R{}, including covariates under the proportional and non-proportional hazards settings.\\

\noindent
{\em Key Words:} survival analysis, hazard rate estimation, non-proportional hazards, CRAN-R.
\end{abstract}

\section{Introduction}

Quantifying patterns in the time to failure and identifying key predictors in prolonging the time to failure are often among the main objectives in analyzing survival data.  The pivotal quantity is generally the hazard rate at time $t$, defined as:
$$h(t) = \lim_{\Delta \to 0} {P(t \le T < t+\Delta \mid T \ge t)}/{\Delta} = {f(t)}/{S(t)},$$
where $S(t) = P(T > t)$ is the survival function and $f(t) = -S'(t)$ is the probability density function of $T$. 
This function provides detailed information on how the hazard of failure changes over the course of a study.  However, $h(t)$ is often difficult to estimate well: there are periods when many failures are observed, causing sharp increases in the the function, and there are times where the number of failures are sparse, providing little information about the true hazard function shape.  

In addition to estimating the hazard rate, quantifying the effects of covariates on time to failure is usually of interest.  Many of the aggregate measures (such as the Kaplan-Meier survival curve \cite{KMest} or Nelson-Aalen cumulative hazard estimator  \cite{NofNA, AofNA}) cannot be estimated simultaneously with covariates.  Instead, the popular Cox proportional hazards model \cite{CoxPH} is often used to determine the effects of covariates and to identify significant predictors of time to failure.  This model is easily implemented in  \R{} using the \makefxn{coxph} function in the {\tt survival} package \cite{coxphsw, survpkg}.  However, the assumption of proportional hazards is not  always satisfied, especially in longer-term studies. These situations require more flexible statistical techniques to estimate the effects of covariates that may vary over time.  In addition, the traditional Cox model does not provide an estimate of the hazard rate, reducing its usefulness in understanding the joint aspects of of the failure process and  effects of predictors.

There are various methodological approaches to estimation of the hazard function, and  a subset of these methodological tools are available as software packages on \cranr \cite{Rref}.  In fact, there are numerous packages available in \R{} that are designed for  semi- or non-parametric estimation of the hazard rate for right-censored survival data.  However, to the average user, the ``best" package can be difficult to select solely based on a package manual. In our experience, it can be  unclear if the implemented methodology is flexible enough to accommodate specific modeling needs, or under what circumstances certain \R{} functions may or may not produce accurate estimates.  In addition, a large barrier to using a package or function is presented when the documentation or implementation does not follow the standard \R{} formatting, making it  onerous to learn new protocols for fitting, summarizing, or plotting survival model results.  

To help navigate this crowded field, this manuscript compares nine of the packages available on \cranr \cite{Rref} which are designed for semi- or non-parametric estimation of the hazard rate for right-censored survival data.  Specifically, we evaluate the availability, speed, and accuracy of the following methodological components:
\begin{enumerate}
	\item Estimation of the hazard rate. \\[-4ex]
	\item Estimation of the effects of one to many covariates. \\[-4ex]
	\item Estimation of  covariate effects under proportional hazards (PH) and non-proportional hazards (NPH) settings.
\end{enumerate}
In addition, we assess if the package and function is ``user-friendly" with the following criteria:
\begin{enumerate}
	\item Adoption of standard \R{} survival formulas for model specification, using the \makefxn{Surv} and \makefxn{formula} functions. \\[-4ex]
	\item Compatibility with the standard \R{} output functions such as \makefxn{print}, \makefxn{plot}, and \makefxn{summary}. \\[-4ex]
	\item Availability of clear and fail-proof documentation that does not require in-depth familiarity with the theory behind the model.
\end{enumerate}

While the last item may seem somewhat controversial to mathematically inclined audiences,  a vast number of everyday users  will not have the requisite familiarity with the methodology behind the software. One way to prevent misinterpretation and misuse of the results is to assure that the documentation and examples are clear and transparent, so  that the basic functions can be used and interpreted without relying on external methodological details from the authors'  manuscript(s). 

This article does not aim to provide an exhaustive review of all \R{}  packages or techniques for analyzing survival data, but instead it aims to  compare -- statistically and practically --  the functions for semi- and non-parametric methods for estimating the hazard rate, possibly in the presence of non-proportional hazard covariate effects. We will inevitably omit a few functions, as we will not be covering cumulative estimators, multistate models, competing risk models, or the incorporation of longitudinal covariates. Our goal is to provide an overview to help others as they try to navigate different package choices by balancing their pros and cons. Ultimately, we hope that the issues we raise and features we comment on will result in  higher overall standards and quality of packages available in \R{} produced by the survival analysis community, and limit the risk of their misuse or nonuse altogether.

The rest of this manuscript is organized as follows: In section 2, we provide a brief overview of the methodology behind each package, as well as its capabilities and code used for model fitting and assessment.  In section 3, we compare the packages using simulated data, including covariates under both the proportional hazards (PH) and non-proportional hazards (NPH) setting. In section 4, we also compare the packages using real data examples for data sets already existing in \R{}, and in section 5 we conclude the manuscript with a discussion.   

\section{Packages and methods}\label{sec:summary}

A list of survival packages with brief summaries can be found on the dedicated \R{} web page \url{http://cran.r-project.org/web/views/Survival.html}. However, many of those packages listed on that page are not regularly maintained, or do not perform all the functions listed in their descriptions.  Based on this and an additional search using Google search terms ``hazard rate estimation" and ``survival analysis", we compiled a list of nine packages that were ultimately included and compared in this manuscript.  The packages included are: \texttt{bayesSurv} \cite{bayessurvR}, \texttt{BGPhazard} \cite{BGPhazardR}, \texttt{bshazard} \cite{bshazardR}, \texttt{DPpackage} \cite{dppkgR}, {\tt MRH} \cite{MRHR}, {\tt muhaz} \cite{muhazR}, {\tt survPresmooth} \cite{survpresmoothR}, {\tt timereg} \cite{timeregPkg}, and {\tt YPmodel} \cite{ypmodelR}.  

These nine packages are designed for semi- or non-parametric estimation of the hazard rate for right-censored survival data, possibly in the presence of proportional and non-proportional hazard covariates.  We briefly summarize the capabilities of each package in Table \ref{tab:PkgCapabs}, as well as indicate whether standard \R{} input and output functions are compatible with the package and whether extra documentation (such as vignettes or manuscripts) are available.  In Table \ref{tab:outputAvail}, we indicate what types of output are available to users (e.g. error bounds on the hazard rate, MCMC chain diagnostics for Bayesian models, etc.), and in Table \ref{tab:PkgBayInput} we indicate which input options users have among the Bayesian functions.  In the rest if this section we provide brief descriptions and references for the methodology and tools behind each of these packages.  The specific aspects that we comment on are:
\begin{itemize}
	\item \underline{\textit{Package and function information}}: Provides a brief description of the purpose of the package and the particular function for fitting survival data, as well as the methodology behind the software.
		\item \underline{\textit{Reliability}}: Overall package performance evaluated on repeated simulated data as well as in analyses of real data sets.  
		\item \underline{\textit{Level of difficulty}}: Overall rating on the ease of package use (``easy", ``moderate", or ``expert").
	\item \underline{\textit{Best use}}: Data and analytic situations for which the package is particularly well-designed. 
	\item \underline{\textit{Other functions available}}: Brief description of any functions that can be applied to the fitted object (e.g. \makefxn{print} or \makefxn{summary}), as well as any other useful pre-processing or post-processing functions.\footnote{Not all functions available in the package are mentioned in this section, as the focus is on those functions that augment the fitted survival or hazard object.}
	\item \underline{\textit{Methodological package limitations}}: Limitations of the package based on the underlying methodology (e.g. the inability to include covariates, estimation issues with small sample sizes, etc).
	\item \underline{\textit{Package implementation limitations}}: Issues and limitations (including internal error messages\footnote{In our discussions of the different packages, we use the term ``internal error message" to denote those error messages that appear when using a function but which are not intended to be interpreted by the user in any manner. In other words, they are not designed as a communication to the user.}) arising from the use of the package or its functions that the users should be made aware of. 
	\item \underline{\textit{Basic code}}: We include the most basic commands required for fitting the models  and obtaining the estimate of the hazard rate and covariate effects (if applicable to that package).  
\end{itemize}

\begin{table}[h!] 
	\center
	\begin{tabular}{|c|l||c|c|c||c|c||c|}
		\hline
		&\multicolumn{1}{|c||}{\multirow{5}{*}{Package}}&\multicolumn{3}{|c||}{Estimate produced}&\multicolumn{2}{|c||}{Standard \R{}}&\multirow{2}{*}{Extra}\\
		\cline{3-5}
		&&&\multicolumn{2}{|c||}{\multirow{2}{*}{Covariate Effects}}&\multicolumn{1}{|c}{}&&\\
		&&Hazard&\multicolumn{1}{|c}{}&&formula&plot and&\multirow{2}{*}{documentation}\\
		&&rate&PH&NPH&for input&summary&\\
		&&&&&&functions&\\
		\hline
		\multirow{4}{*}{Bayesian}&{\tt bayesSurv}&x&-$^\dagger$&x&x&-&-\\
		&{\tt BGPhazard}&x&-&-&-&-&V\\
		&{\tt DPpackage}&x&-$^\dagger$&x&x$^\ddagger$&x&M$^\S$\cite{DPpkgJSS}\\
		&{\tt MRH}&x&x&x&x&x&V\\
		\hline
		\multirow{5}{*}{Frequentist}&{\tt bshazard}&x&x&-&x&x&M\cite{bshazardMan}\\
		&{\tt muhaz}&x&-&-&-&x&-\\
		&{\tt survPresmooth}&x&-&-&-&-&M\cite{presmoothJSS}\\
		&{\tt timereg}&x&x&x&x&x&M$^\S$\cite{ScheikeCR}, B\cite{timereg}\\
		&{\tt YPmodel}*&-&-&only 1&-&x**&O\\
		\hline
	\end{tabular}
	\caption{\footnotesize Comparison table of packages' capabilities, noting which packages provide functions that can estimate the hazard rate, the effects of covariates under the proportional hazards (PH) and non-proportional hazards (NPH) assumptions, and whether standard \R{} input and output functions are available.  In the table, `x' denotes that the package includes functions that have that specific capability, and `-' denotes that the package does not provide any such function.  The ``Standard \R{} functions" column refers to the conversion of the standard \makefxn{plot}, \makefxn{print} and \makefxn{summary} functions to be used on fitted objects from the package.  For example, the {\tt foo} package may contain \makefxn{plot.foo}, allowing the user to produce unique plots only available on a fitted object created using a function from the {\tt foo} package via the command \texttt{plot(foo.fitted.object)}.  See \cite{createRpkg} for more on this topic.  The ``Extra documentation" column denotes whether the library has extra documentation on how to use the package and/or the function presented in the manuscript through a vignette (``V"), manuscript (``M"), book (``B"), or other (``O").  \newline\newline
	*Although the {\tt YPmodel} package does not offer a function that estimates the hazard rate, we include it because it provides an estimate of the log-ratio between two non-proportional hazards.  \newline\newline
	**The standard \R{} plot and summary functions are nominally available for this package, but they do not actually perform the standard plotting or summary actions.  All three functions provide the same output: multiple plots and summaries and an internal error message that stops the routine(s) mid-way through. \newline\newline
	$^\dagger$ Estimated covariate effects are available, but they are not interpreted in the manner of the traditional Cox PH model.\newline\newline
	$^\ddagger$ The formula for the \lddp{} deviates from the traditional formula set-up.  See Section \ref{sec:dppkg} for details.\newline\newline
	$^\S$ Extra documentation is available for the package, but it does not contain code or examples on using the survival function mentioned in this manuscript.}
	\label{tab:PkgCapabs} 
\end{table}

\begin{table}[h!] 
	\center
	\begin{tabular}{|c|l||c|c|c|c||c||c|c|}
		\hline
		&\multicolumn{1}{|c||}{\multirow{4}{*}{Package}}&\multicolumn{4}{|c||}{\multirow{2}{*}{Estimate (Error Bounds)}}&\multirow{2}{*}{Model}&\multirow{2}{*}{MCMC}&\multirow{2}{*}{Chain}\\
		&&&&&&&&\\
		\cline{3-6}
		&&Hazard&Survival&Cumulative&Covariate&Selection&\multirow{2}{*}{Chains}&\multirow{2}{*}{Diags}\\
		&&Rate&Curve&Hazard&Effects&Criteria&&\\
		\hline
		\multirow{4}{*}{Bayesian}&{\tt bayesSurv}&x* (x*)&x* (x*)&x* (x*)&x**$^\dagger$ (x**)&-&x**&-\\
		&{\tt BGPhazard}&-&x$^\ding{64}$ (-)&-&&-&x&x*\\
		&{\tt DPpackage}&x (x)&x (x)&-&x (x*)&CPO&x**$^\S$&x*\\
		&{\tt MRH}&x (x)&x (x)&x (x)&x (x)&AIC, DIC, BIC&x**&x*\\
		\hline
		\multirow{5}{*}{Frequentist}&{\tt bshazard}&x (x)&-&-&-&-&&\\
		&{\tt muhaz}&x (-)&-&-&&-&&\\
		&{\tt survPresmooth}&x (-)&x$^\ddagger$&x$^\ddagger$&&-&&\\
		&{\tt timereg}&x* (-)&-&-&x (x)&param tests&&\\
		&{\tt YPmodel}&&&&x (x)&-&&\\
		\hline
	\end{tabular}
	\caption{\footnotesize Comparison table of output available directly from the fitted model (denoted with ``x"), through a post-processing function provided by the package (``x*"), or in an output folder (``x**").  A ``-" denotes that the output feature is not available, and an empty cell denotes that the feature is not applicable to that function. The table contains information on whether hazard rate, survival curve, cumulative hazard, and covariate effect estimates and error bounds are accessible to the user (not including graphics), as well as model selection criteria (includes AIC, BIC, DIC, CPO, and parameter significance testing), and if the MCMC chains and/or diagnostics are available (only applicable to Bayesian models). 
	\newline\newline
		$^\dagger$ The estimated covariate effects can be calculated from their posterior distributions by reading in the corresponding text file of the MCMC chains.  However, there are no post-processing functions to carry this calculation out directly.\newline\newline
		$^\ding{64}$ The survival curves are calculated for each stored MCMC iteration and are returned in the fitted {\tt BGPhazard} object.  A point estimate and credible intervals can be calculated by the user via the the chain of the survival curves (through the mean or median of the posterior distribution).\newline\newline
	$^\S$ The MCMC chains are available, but require FORTRAN code to access and process them.\newline\newline
	$^\ddagger$ Estimates of the survival curve and cumulative hazard curve can be obtained using \presmooth{} (the same function used to estimate the hazard rate), but by setting the ``estimand" option to ``S" or ``H". (See Section \ref{sec:presmooth} for details.)}
	\label{tab:outputAvail} 
\end{table}

\begin{table}[h!] 
	\center
	\begin{tabular}{|l||cc|ccc|c|c|}
		\hline
		\multicolumn{1}{|c||}{\multirow{3}{*}{Package}}&\multicolumn{2}{|c|}{Parameter Values}&\multicolumn{3}{|c|}{MCMC Sampling}&Continue&Multiple\\
		\cline{2-6}
		&Initialize&Set&\multirow{2}{*}{Thin}&\multirow{2}{*}{Burn-in}&Total&chain&chains\\
		&prior&hyperprior&&&iterations&&\\
		\hline
		{\tt bayesSurv}&x&x*$^\dagger$&x&x&x&-&-\\
		{\tt BGPhazard}&x&x&x&x&x&-&-\\
		{\tt DPpackage}&x*&x*&x*&x*&x*&x&x\\
		{\tt MRH}&x&x&x&x&x&x&x\\
		\hline
	\end{tabular}
	\caption{\footnotesize Comparison of the input options for the Bayesian packages.  An `x' denotes that the user has the option to specify the input, but a default is available, and an `x*' denotes that the user \textit{must} specify this input.  A `-' denotes an option that cannot by specified by the user for that function.  The ``continue chain" column denotes the option for users to continue the chain after the function has finished running (useful, for example, in the instance of a chain that has not yet converged but the maximum number of iterations has been reached).  The ``multiple chains" column denotes the option for the user to specify that multiple chains will be run for the same data set, and that the initialized values of the sampled parameters should vary (useful, for example, in the instance of performing a Gelman-Rubin diagnostic convergence test \cite{gelmanrubin} on multiple chains run on the same data set).\newline\newline
	$^\dagger$ The package has default values for the hyperpriors, but in most practical instances, the user must specify the hyperpriors themselves for the routine to work (this occurs, for example, when the model includes covariates).}
	\label{tab:PkgBayInput} 
\end{table}

\subsection{{\tt bayesSurv} package} 
\begin{itemize}
\item \underline{\textit{Package and function information}}:  The \texttt{bayesSurv} package is designed for Bayesian survival regression that can accommodate flexible error and random effect structures.  The function we examine in this paper is \baysurv{}, which is a Bayesian accelerated failure time (AFT) survival model that can accommodate clustered data, and is of the form: 
$$\log(T_{ij}) = X_{ij}'\beta + z_{ij}'b_i+\epsilon_{ij}, i = 1, \dots, n, j = 1, \dots, n_i.$$
Here, the distribution of $\epsilon_{ij}$ is specified as a normal mixture with an unknown number of components as in \cite{bayessurvRichardson} and the random effect $b_i$ is normally distributed.  (See \cite{bayessurvKomarekthesis} or \cite{bayessurvKomarek} for a more detailed description of prior assumptions.) Fixed and random effects for the model are estimated using classical Bayesian mixed model approaches (such as that found in \cite{BayesianMixedMods}), and the hazard rate is approximated based on a mixture of normal distributions.  For the purposes of this paper, we solely investigate simple survival models without any random effects, implemented via the \baysurv{} function. We do not explore this package in its full capacity, but only investigate simple right-censored survival models.
	\item \underline{\textit{Reliability}}: Moderate to excellent, depending on the type of survival data analyzed.
	\item \underline{\textit{Level of difficulty}}: Expert.  
	\item \underline{\textit{Best use}}:   The {\tt bayesSurv} package itself presents many advantages, as it contains functions to analyze more sophisticated types of data that require statistical methods generally not found in standard \R{} packages, such as bivariate and/or clustered data survival analysis.  However, for simpler analyses such as those presented in this manuscript, there may not be much advantage to learning the  \baysurv{} function  when more straightforward yet equally reliable alternatives are available (see Discussion).    
\item \underline{\textit{Functions available in the {\tt bayesSurv} package (relevant to this manuscript)}}: 
\begin{itemize}
	\item \baysurv{}:  Fits the survival model by sampling from the posterior distribution.
	\item \makefxn{predictive}: Provides a sample of hazard rates calculated using the parameter chains created using \baysurv{} for specified subgroups (i.e. treated vs untreated subjects).
	\item \makefxn{give.summary}: Summarizes each parameter in the chains obtained using MCMC. 
\end{itemize} 
	\item \underline{\textit{Methodological package limitations}}: 
		\begin{itemize}
			\item \textbf{Only AFT modeling for covariate effects}.  While AFT models allow for non-proportional hazards, the hazards cannot cross, limiting the scenarios under which this model is appropriate.  In addition, the covariate effects in AFT models are not interpreted in the same manner as the Cox PH covariate effects, necessitating careful interpretation of the results.
			\item \textbf{Longer run-time.}  Takes longer to converge than some frequentist methods due to MCMC sampling.
		\end{itemize}
	\item \underline{\textit{Package implementation limitations}}:
		\begin{itemize}
			\item \textbf{Complicated examples.}  The function examples are not included in the manual page, and instead example code must be obtained from the author's website or from a folder in the downloaded package.  Once the code is obtained, the examples are complex and require extra coding by the user.\footnote{The code contains minimal comments as to the exact purpose of each step, requires the user to create multiple output folders and files (some of which do not match up with each other), and has multiple steps that could be consolidated for easier understanding of the use of each function.} This lack of clarity in the examples is surprising, as the model input and output parameters are very well-documented in the manual page, with details provided as to the purpose of each user option.
			\item \textbf{Posterior distribution of the hazard rate(s) unavailable.} While there are post-processing functions that allow the user to calculate any desired quantile of the hazard rate (or survival function and cumulative hazard function), the samples of the posterior distributions for the hazard rates are unavailable.  These samples would be particularly useful for users who want to calculate the estimated log-hazard ratio between two hazard rates, as well as the credible interval for the log-hazard ratio.  An estimate can be obtained by calculating the log of the ratio of the medians or the means of the posterior distributions, error bounds on the estimate must be approximated.\footnote{In this manuscript, we approximate the point-wise variance of each hazard rate using the upper and lower credible interval bounds, which we then used to approximate the point-wise variance (and therefore 95\% confidence intervals) of the log-hazard ratio.\label{footnote:boundsapproxBS}}
		\end{itemize}
 
\hrule 
\item \underline{\textit{Basic code}}: Basic commands and detailed comments for fitting the {\tt bayesSurv} model and extracting the estimates and their uncertainty are:
{\small \begin{verbatim}
	# Set the values of the parameters in the hyper priors, and initialized the other parameters.
	# This can be done in the function directly but is shown separately for the purpose of clarity
	prior.values = list(kmax = 30, k.prior = 'poisson', poisson.k = 5, dirichlet.w = 1, mean.mu = 0, 
                       var.mu = 5^2, shape.invsig2 = 2.0, shape.hyper.invsig2 = 0.2, 
                       rate.hyper.invsig2 = 0.1, pi.split = c(1, rep(0.5, kmax - 2), 0), 
                       pi.birth = c(1, rep(0.5, kmax - 2), 0), Eb0.depend.mix = FALSE)
	initialized.values = list(iter = 0,  
	                    mixture = c(1, 1, rep(0, prior$kmax - 1), rep(0, prior$kmax), rep(1, prior$kmax)), 
                    beta = 0, D = NULL, b = NULL, r = rep(1, 1000), 
                    otherp = rgamma(1, shape = prior$shape.hyper.invsig2, rate = prior$rate.hyper.invsig2), 
                    u = c(runif(1), 0, 0, runif(3*(prior$kmax - 1))))
	prior.betaavalues = list(mean.prior = 0, var.prior = 25)
	
	# Set the MCMC specifications.  This can be done in the function directly,
	# but is declared here separately for the purpose of clarity.
	mcmcinfo = list(niter = 10000, nthin = 10, nburn = 2000, nnoadapt = 0, nwrite = 100)
	
	# Sample the parameters from the posterior distribution via MCMC.  
	# Results are stored in a specified directory (in this example named "bayessurvoutput")  
	bayesSurv.fit = bayessurvreg1(Surv(time, delta) ~ covariates, data = data, 
                                init = initialized.values, prior = prior.values, 
                                prior.beta = prior.betavalues, nsimul = mcmcinfo, dir = 'bayessurvoutput')
					
	# Calculate the hazard rate for a predicted data set at specified time points.
	# The predicted data set must include all covariates used in the model fit, and 
	# must be named in the same manner as the original covariate set.
	fithr = predictive(Surv(1, 1) ~ covariates, data = predicted.dataset, quantile = 0.5, skip = 0, by = 1,
                      predict = list(Et=FALSE, t=FALSE, Surv=FALSE, hazard=TRUE, cum.hazard=FALSE), 
                      store = list(Et=FALSE, t = FALSE, Surv = FALSE, hazard = FALSE, cum.hazard=FALSE), 
                      dir = 'bayessurvoutput', grid = predicted.timepoints)
	
\end{verbatim}}
\end{itemize}
\hrule

\subsection{{\tt BGPhazard} package} 
\begin{itemize}
\item \underline{\textit{Package and function information}}: This package estimates the hazard rate using Markov Beta and Gamma processes as described in \cite{nieto} and \cite{BGPhazard}.  The function we examine in this manuscript is \bgp, which employs a Bayesian, non-parametric, Markov Beta model to examine the failure process of discrete survival times (i.e. survival times that are located at equally spaced, distinct periods of time). The hazard increments are modeled through a Beta process, with prior correlation between adjacent increments introduced via a latent Poisson process.  This package is based on methodological developments found in \cite{nieto}. 
	\item \underline{\textit{Level of difficulty}}: Expert.
	\item \underline{\textit{Reliability}}: Moderate, except in instances where the number of discrete time points is dramatically larger than the number of subjects, in which case this method is not optimal as the estimates become too variable.
	\item \underline{\textit{Best use}}:  In instances where survival times are naturally discrete and a sufficient number of failures are observed at each discrete time point.  In addition, because the method and software allow the user to control the degree of dependence between the hazard rate estimates at each time point (through the prior), which may be of interest in some instances.  
\item \underline{\textit{Functions available in the {\tt BGPhazard} package}}:\footnote{The {\tt BGPhazard} package also contains functions for estimating the hazard rate using a Gamma process (we were unable to use this function without receiving warnings messages or internal errors) and for the analysis of survival data with time-dependent covariates (which are not applicable to this work and are therefore not included in the manuscript).}
\begin{itemize}
	\item \bgp{}:  Fits the survival model.
	\item \makefxn{BePlotDiag}: Used to examine chain behavior (i.e. trace, autocorrelation).
	\item \makefxn{BePloth}: Used to plot the estimated hazard rate and credible bands.
\end{itemize} 
	\item \underline{\textit{Methodological package limitations}}: 
		\begin{itemize}
			\item \textbf{Discrete survival times required.} This may require the user to round or convert the observed survival times.\footnote{Zeroes are not allowed in the model, possibly requiring the user to round, rescale, or convert the survival times using some other method.  Note that simply re-scaling the times may introduce many more time points than subjects, causing the estimation routine to become unstable.}
			\item \textbf{Cannot accommodate covariates.}
			\item \textbf{Unreliable estimation of the hazard rate when the sample size is small and the number of time points is large.} As each discrete time point requires a certain number of parameters, this method can have trouble converging when the number of time points is large compared to the number of subjects in the data set.\footnote{It is possible to mitigate this problem by increasing the level of correlation between the bins of the hazard through the prior, but may not always be successful if the sample size is too small.}
			\item \textbf{Long run-time.} Due to the MCMC sampling, this Bayesian function takes longer to run than the frequentist methods we reviewed.
		\end{itemize}
	\item \underline{\textit{Package implementation limitations}}:
		\begin{itemize}
			\item \textbf{Not enough documentation.}  Given the advanced level of statistics behind these methods, the help pages of the function and package do not contain much explanation on each of the input and output function components.\footnote{For example, the \bgp{} function has an option to adjust the parameter ``alpha", and the entire description of this component is ``Nonnegative vector. Small entries are recommended in order to specify a non-informative prior distribution." }  Even for a well-informed user who has read the manuscripts and vignette, more details on the help page would be useful as reminders as to which components of the function are associated with the different pieces of the estimation routine.  Documentation is also sparse for post-processing of the MCMC chains and on how to use other aspects of the fitted object.
			\item \textbf{Difficulty in obtaining the hazard rate estimate.}  Plots of the hazard rate and survival curve can be created by the user,\footnote{Plots are created through the \makefxn{BePloth} function.} but the point estimates are not directly available in the fitted model or through any post-processing functions, and locating the posterior distribution of the hazard rate also poses difficulties.\footnote{The posterior distribution of the hazard rate is contained in the matrix of the MCMC chains (located in the ``summary" component of the fitted object).  However, the MCMC matrix does not have labeled columns, and it is difficult to determine which parameters are associated with the columns of the matrix of chains.  The manuscript and theoretical model must be understood before the matrix can be used for summary of the parameters and the hazard rate. No documentation on obtaining the posterior distribution of the hazard rate or the corresponding point estimate is provided in the user manual or the vignette.}
		\end{itemize}
\hrule
\item \underline{\textit{Basic code}}: Basic commands for fitting the BGP model and extracting the estimates and their uncertainty are:
{\small \begin{verbatim}
	# Fit the model.  Time must be discrete, integer values, 
	# which may require rounding to the nearest whole number.  
	# May need to use the ceiling function if there are survival times less than 0.5.
	BGP.fitPH = BeMRes(round(dataset$time), dataset$delta)
	
	# Obtain the hazard rate estimate (in this example, it is calculated as the
	# mean of the marginal posterior distribution for the hazard rate at each
	# time point)
	BGPhazrate = rowMeans(BGP.fitPH$summary[1:max(BGP.fitPH$tao),])
	
	# Uncertainty estimates can be calculated using the chains.  
	# For example, the 95% central credible interval can be calculated as:
	BGPbounds = sapply(1:max(BGP.fitPH$tao), function(x) 
		quantile(BGP.fitPH$summary[x,], probs = c(.025, .975)))
\end{verbatim}}
\end{itemize}
\hrule

 \subsection{{\tt bshazard} package}\label{sec:bshazard}
 \begin{itemize}
	\item \underline{\textit{Package and function information}}: This package estimates the hazard function using B-splines from the perspective of generalized linear models (i.e. assuming a Poisson distribution; \cite{bshazard1, bshazard2}).\footnote{The \bshaz{} function can accommodate both left-truncated and right-censored data, however we only examine the case of right-censored data in this manuscript.}  Users may specify the number of bins the hazard rate should be estimated over, the number of knots and the degree of the B-splines, and the value for the estimation of the smoothing parameter.  Estimation is done via the \bshaz{} function.
	\item \underline{\textit{Reliability}}: Excellent.
	\item \underline{\textit{Best use}}: Highly recommended under the proportional hazards setting, particularly in instances that do not typically cause problems with spline estimation (see Footnote \ref{footnote:smoothprobs}).	
	\item \underline{\textit{Level of difficulty}}: Easy.
	\item \underline{\textit{Functions available in the {\tt bshazard} package}}:
	\begin{itemize}
		\item \bshaz{}: Fits the survival model.
		\item \makefxn{plot}, \makefxn{print}, \makefxn{summary}:  standard \R{} functions for assessing the \bshaz{} fitted model.  
	\end{itemize}
	\item \underline{\textit{Methodological package limitations}}: 
		\begin{itemize}
			\item \textbf{Standard possible issues with spline estimation}.\footnote{Examples of issues with spline and kernel smoothers include possible estimation inaccuracy near the boundaries of the study, and the trade-off between an increased variance and capturing all the peaks and valleys of the true hazard rate.  In addition, for kernel smoothing, the choice of the kernel may heavily influence the shape of the hazard rate. \label{footnote:smoothprobs}}
		\end{itemize}
	\item \underline{\textit{Package implementation limitations}}: 
		\begin{itemize}
			\item \textbf{No error bounds or variance estimates provided for covariate effects.}
			\item \textbf{Average hazard rate returned to user.}  The hazard rate returned to the user is not the standard baseline hazard rate, but is the hazard rate for a subject with the average measurement of each coefficient added to the model.  This also poses issues in estimation of the 95\% error bounds.\footnote{Conversion of the \bshaz{} hazard rate estimate to the baseline hazard rate estimate can be done by multiplying the estimated hazard rate for the average subject by $\exp(-\bar{X}'\hat{\beta})$, where $\bar{X}$ is the average user covariate values and $\hat{beta}$ is the estimated covariate effect.  The upper and lower interval bounds can be calculated for the baseline hazard in a similar manner, however, these converted bounds do not take into account the variation of $\hat{\beta}$ (a number which is not provided by the model) and so the baseline intervals are likely narrower than if all information were provided. \label{footnote:bshazvar}}
			\item \textbf{Error messages in some data analyses.} See Section \ref{sec:PHexample}.
		\end{itemize}
\hrule
\item \underline{\textit{Basic code}}:
This routine accepts zero to many covariates under the proportional hazards assumption, and uses the standard \R{} input formula for model creation.  The general code for fitting the proportional hazards \bshaz{} model and obtaining the estimated parameters and variances is:
{\small\begin{verbatim}
	# Fit the model using standard R formula for input
	bshazard.fitPH = bshazard(Surv(time, delta) ~ covariates, data = data)
	
	# Obtain the estimated hazard rate (for the average group), 
	# and error bounds, as well as estimated covariate effects
	# (no error bound provided for covariate effects).
	bshazrate = bshazard.fitPH$hazard
	bshazrate.lowerCI = bshazard.fitPH$lower.ci
	bshazrate.upperCI = bshazard.fitPH$upper.ci
	bscovars = bshazard.fitPH$coeff
	
	# Obtain the  estimated hazard rate for the baseline group
	bshazrate.base = bshazrate*exp(bscovars*bshazard.fitPH$covariate.value)
\end{verbatim}}
The estimated hazard rate and covariate effects can also easily be obtained using the \makefxn{summary} function on the fitted object.
\end{itemize}
\hrule
 \subsection{{\tt DPpackage} package} \label{sec:dppkg}
 \begin{itemize}
	\item \underline{\textit{Package and function information}}: This package contains functions to perform estimation and inference for many complex Bayesian non-parametric and semi-parametric models based on the Dirichlet Process prior.  The function we examine in this manuscript is \lddp{}, which generates a posterior density sample for a Linear Dependent Dirichlet Process mixture of survival models.  This model is an extension of the  ANOVA Dependent Dirichlet Process initially presented in \cite{IorioAnova}, and performs survival regression based on a Dirichlet Process prior \cite{Iorio}. In the model, the set of the random probability distributions are dependent in an ANOVA-type fashion. If $\{F_x, x \in X\}$ is the set of random distributions indexed by the categorical covariates $x = (x_1,\dots,x_p)$, defined as
$$F_x = \sum_{h = 1}^\infty p_h\delta(\theta_{xh}),\text{ for each }x \in X,$$
such that $\sum_{h = 1}^\infty p_h = 1,$ and $\delta(y)$ represents a point mass at $y,$ then dependence is introduced by modeling the locations $\theta_{xh}$ through the covariates (see \cite{IorioAnova} and \cite{Iorio} for details).  This method allows all group-specific hazards to be modeled non-proportionally and can accommodate continuous covariates.\footnote{The \lddp{} function can also accommodate interval-censored data, although that is not examined in this work.}
	\item \underline{\textit{Reliability}}: Excellent when the number of covariates is small.
	\item \underline{\textit{Level of difficulty}}: Expert.
	\item \underline{\textit{Best use}}: This package is ideal for researchers interested in implementing models using the Dirichlet Process, and contains many useful functions that are hard to find elsewhere on \R{} and would require intensive coding.  In a simple survival analysis setting, this package is difficult for those only familiar with the interpretation of covariates in the Cox PH model or for those who only need to implement a straightforward survival model.  
	\item \underline{\textit{Functions available in the {\tt DPpackage} package}}:\footnote{{\tt DPpackage} includes two other survival analysis functions based on the DP prior:  \makefxn{LDTFPsurvival}, which uses the tail-free process to model the survival curve, or \makefxn{DPsurvint}, which is an AFT model based on the Dirichlet Process. We omit them from this manuscript as they have similar implementation schemes to \lddp{}.}\footnote{{\tt DPpackage} also contains an extensive suite of other functions, including models that examine Dirichlet Processes, Dependent Dirichlet Processes, Dependent Poisson Dirichlet Processes, Hierarchical Dirichlet Processes, Polya Trees, Linear Dependent Tailfree Processes, Mixtures of Triangular distributions, Random Bernstein polynomials priors and Dependent Bernstein Polynomials, Penalized B-Splines, semiparametric models for marginal and conditional density estimation, ROC curve analysis, binary regression models, generalized linear mixed models, IRT type models, and generalized additive models. These aspects of the package are not examined as they are not applicable to the work in this manuscript.} 
		\begin{itemize}
			\item \lddp{}: Estimates the survival function using the Dependent Dirichlet Process.  
			\item \makefxn{plot}, \makefxn{print}, \makefxn{summary}: standard \R{} functions for assessing the \lddp{} fitted model.
		\end{itemize} 
	\item \underline{\textit{Methodological package limitations}}: 
		\begin{itemize}
			\item \textbf{Interpretation of covariate effects.}  The effects of the covariates on the time to failure can be interpreted in the same manner as ANOVA parameters, but cannot easily be converted to the familiar Cox PH model estimates.  In addition, there is no way to impose proportionality between hazard rates with this model.
			\item \textbf{Limitation on the number of covariates.} Use of a Dirichlet Process prior may require that only a small number of covariates be included in the model, or estimation issues occur and the function tends to crash (see Section \ref{sec:nphex}).
			\item \textbf{Longer run-time.}  Takes longer to converge than some frequentist methods due to MCMC sampling.
		\end{itemize}
	\item \underline{\textit{Package implementation limitations}}: 
		\begin{itemize}
			\item \textbf{Accessing MCMC chains is difficult.} FORTAN must be used to read or analyze the output files, making it difficult for the average user to access the chains.\footnote{Similar to the {\tt bayesSurv} package, the posterior distributions of the hazard rates (or survival curves and cumulative hazard functions) are unavailable to the user.  This becomes particularly difficult if a user would like to calculate the log-hazard ratio and credible interval bounds.  If this is the case, these calculations must be approximated (see Footnote \ref{footnote:boundsapproxBS}).} 
			\item \textbf{Post-processing function needed.} The user may predict a hazard rate for one or more subject profiles, however, this must be done using \lddp.  If the user later would like a different profile, the routine must be run again to obtain this information.  A post-processing function that allows the user to access the existing chains to calculate predicted hazard rates would be useful. 
			\item \textbf{More documentation needed for values contained in fitted object.} The ``Values" section of the manual page contains information on how to re-start the chain, but no information is available on the components of the fitted survival object.
			\item \textbf{Non-standard model input formula.}  The survival formula for creating the fitted model has a some glitches that can be tricky to navigate.\footnote{The data must be entered as interval censored data, and can only be entered using the \makefxn{cbind} method, not the \makefxn{Surv} method.   In addition, subjects that are right-censored must have end times entered as ``-999", which is converted inside the function to a very large number (meant to represent infinity).  However, this requirement for converting the input is not documented in the manual page.}
			\item \textbf{Error message needed.} When too many covariates are included in the model, the program tends to crash.  Because there is no error message to communicate this issue to the user, it is difficult to figure out the exact source of the problem.
		\end{itemize}
\hrule
\item \underline{\textit{Basic code}}:
The general code for fitting the \lddp{} model and obtaining the estimated parameters and variances is below.  For expository purposes, we only give code for one covariate (treatment), but other covariates can be included.
{\small \begin{verbatim}
# Set the priors and specification for the MCMC sampling.
prior = list(a0=10, b0=1, nu=4, m0=rep(0,3), S0=diag(100,3), psiinv=diag(10,3),
                tau1=5, taus1=5, taus2=2)
mcmc = list(nburn=500, nsave=5000, nskip=10, ndisplay=1000)

# Convert the right-censored data to a start time and end time 
# so that the interval censored approach can be used.
# Those that are right-censored must have end times set equal to -999. 
starttime = endtime = dataset$time
endtime[dataset$censored == 0] = -999

# Fit the survival model.
# The ``zpred" option is used to specify which predicted hazard rates the 
# user would like returned to them.  In this example, one hazard rate will
# be the baseline hazard rate, and the other hazard rate will be the 
# treatment hazard rate.  
# The ``grid" option are the time points over which the user would
# like the hazard predicted for them.
fit = LDDPsurvival(cbind(starttime, endtime) ~ treatment,
                                prior = prior, mcmc = mcmc, zpred = rbind(c(1,0), c(1,1)),
                                grid = time.points, status = TRUE)

# Extract the baseline and treatment hazard rates from the fitted object.
hazard.baseline = fit$hazp.m[1,]
hazard.treatment = fit.complex$hazp.m[2,]
\end{verbatim}}
\end{itemize}
\hrule

\subsection{{\tt MRH} package}
\begin{itemize}
\item \underline{\textit{Package and function information}}: The {\tt MRH} package contains functions that are used to jointly estimate the hazard rate and the effects of covariates on the failure time for right-censored data.  Covariates can be included under both the proportional and non-proportional hazards setting.  The function we examine in this manuscript is \emrh{}, which a Bayesian semi-parametric method that employs the multi-resolution hazard Polya tree methodology (see \cite{Bouman,Bouman2,Dukic}).  This methodology splits the hazard rate into $2^M$ bins, and estimates a constant hazard rate within each bin using a tree-like prior structure.  If $h_{base}(T)$ denotes the baseline hazard rate up to time $T$, $S_{base}(T)$ represents the survival function up to time $T$, and $\delta$ is the censoring variable, the likelihood function for the \emrh{} model is written as:
\begin{align}
	L(\mathbf{T \mid \vec \beta, X, T}) = \prod_{\ell = 1}^\mathcal{L}\prod_{i \in \mathcal{S}_\ell}
			\left[h_{base, \ell}(T_{i, \ell})e^{X_{i, \ell}'\vec \beta}\right]^{\delta_{i, \ell}}S_{base, \ell}(T_{i, \ell})^{e^{X_{i, \ell}'\vec \beta}} \label{eqn:MRHlikelihood}.
\end{align}
Covariates are included under the proportional hazards assumption through the  covariate matrix $X$, and the covariate effects are modeled through the parameter vector $\vec \beta$. 
Covariates are included under the non-proportional hazards assumption through separate hazard rates for each of the $\ell$, $\ell = 1, \dots, \mathcal{L}$ non-proportional covariate strata (denoted as $\mathcal{S}_\ell$).  The estimated log-hazard ratio of these time-varying effects are then calculated as the log of the ratio of the hazard rate of stratum $\ell$ to stratum 1 (if stratum 1 represents the baseline group).  The hazard rate for each stratum $\ell$ is given a Polya-tree-like, wavelet-based, multi-resolution prior.  (For more details, see  \cite{Bouman,Bouman2,Dukic,Dignam}.)
 	\item \underline{\textit{Reliability}}: Excellent.
	\item \underline{\textit{Level of difficulty}}: Easy to moderate based on user's knowledge of Bayesian methods.
	\item \underline{\textit{Best use}}:  Highly recommended for estimation under both the PH and NPH settings, particularly for studies with periods of sparse observed failures.\footnote{The \emrh{} function also allows for bins with similar hazard rates to be combined, or ``pruned" (not covered in this manuscript, see {\tt MRH} vignette and \cite{Yprune} for details), which increases the efficiency of the sampling algorithm and can provide a smoother estimated hazard rate.}  While reliable, for simple survival models on large data sets, other options may be more straightforward (see Discussion).
\item \underline{\textit{Functions available in the {\tt MRH} package}}: 
 \begin{itemize}
 	\item \emrh{}: Fits the survival model.
 	\item \makefxn{plot}, \makefxn{summary}: Standard \R{} functions for assessing the fitted model.
	\item \makefxn{AnalyzeMultiple}: Analyzes convergence and produces output information for the MCMC chain(s).
 \end{itemize}
	\item \underline{\textit{Methodological package limitations}}: 
		\begin{itemize}
			\item \textbf{Hazard rate estimated over a fixed number of bins.} The number of bins must be a factor of 2 and must be of equal length (see code below).
			\item \textbf{NPH covariates must be factors.} A hazard rate is estimated for each of the NPH strata, so any NPH covariates included in the model must be factors (or continuous variables categorized).
			\item \textbf{Longer run-time.}  Takes longer to converge than some frequentist methods due to MCMC sampling.
		\end{itemize}
	\item \underline{\textit{Package implementation limitations}}: 
		\begin{itemize}
			\item \textbf{Sampling error.} In rare instances, within the first few iterations of the MCMC routine, the function used to sample the parameters becomes numerically unstable and crashes, requiring the routine to be called again.
		\end{itemize}
\hrule
\item \underline{\textit{Basic code}}:
The general code for fitting the {\tt MRH} model and obtaining the estimated parameters and credible intervals (under both PH and NPH circumstances) is:
{\small \begin{verbatim}
# Obtain the fitted model values for the PH model for a model with 8 equally spaced bins
MRH.fitPH = estimateMRH(Surv(time, delta) ~ covariatesPH, data = data, M = 3)
# Obtain the fitted model values for the NPH model for a model with 8 equally spaced bins
MRH.fitNPH = estimateMRH(Surv(time, delta) ~ covariatesPH + nph(covariateNPH), data = data, M = 3)

# Obtain the estimated hazard rate, 
# covariate effects and error bounds directly from the model.
# Similar code can be used on the NPH models.
MRHhazrate = MRH.fitPH$hazrate
MRHcovars = MRH.fitPH$beta
\end{verbatim}}
In the above code, $M$ controls the levels of the prior tree, resulting in $2^M$ estimated hazard increments.  The estimated hazard rate and covariate effects can also easily be obtained through the \makefxn{summary} function on the fitted object or on the MCMC chains. 
\end{itemize}
\hrule
\subsection{{\tt muhaz} package}
\begin{itemize}
	\item \underline{\textit{Package and function information}}: The \muhaz{} function in the {\tt muhaz} package produces a smooth estimate of the hazard function for right-censored data using kernel-based methods. Options for estimation include three types of bandwidth functions, three types of boundary correction, and four shapes for the kernel function, employing global and local bandwidth selection algorithms \cite{muhaz1}, and nearest neighbor bandwidth formulation \cite{muhaz2}.\footnote{In the \muhaz{} function, users may provide specifications for the bandwidth, the number of estimation points in the grid, the choice of the kernel function (possible values are ``epanechnikov" (default), ``rectangle", ``biquadratic", and ``triquadratic" ), and the time range over which the hazard rate function is estimated (default is the first and last uncensored failure time).}  The underlying theoretical properties of the estimator used in this package can be found in \cite{muhaz3}.
	\item \underline{\textit{Reliability}}: Moderate to excellent.
	\item \underline{\textit{Level of difficulty}}:  Easy.
	\item \underline{\textit{Best use}}:  Recommended for initial exploratory analyses of the shape of the hazard rate when no covariates are included, as it is very simple and quick to learn.  For a more thorough examination of a data set, other functions are more appropriate.
	\item \underline{\textit{Functions available in the {\tt muhaz} package}}:\footnote{The {\tt muhaz} package also contains the hazard rate estimation function \makefxn{pehaz}, which calculates the piecewise exponential hazard rate estimates.  However, in using this function we regularly encountered estimation issues (due to periods of time with no observed failures) and therefore it was not included in this manuscript.}
	\begin{itemize}
		\item \muhaz{}: Fits the survival model.
		\item \makefxn{plot}, \makefxn{print}, \makefxn{summary}: Standard \R{} functions for assessing the fitted model.
	\end{itemize}
	\item \underline{\textit{Methodological package limitations}}: 
		\begin{itemize}
			\item \textbf{Cannot accommodate covariates.}
			\item \textbf{No uncertainty bounds for the hazard rate.}\footnote{The \muhaz{} fitted object contains the integrated mean square error (IMSE) of the hazard rate, but this would likely need to be de-cumulated and smoothed to be applicable to the hazard rate.  (Instructions for this are not given in the documentation.)}
			\item \textbf{Standard potential issues with kernel smoothing estimation.} See Footnote \ref{footnote:smoothprobs}.
		\end{itemize}
	\item \underline{\textit{Package implementation limitations}}: 
		\begin{itemize}
			\item \textbf{Non-standard model formulation.} The standard \R{} \makefxn{Surv} and \makefxn{formula} functions are not used for input, although this is not an issue as no covariates are included in these models.
		\end{itemize}
\hrule
\item \underline{\textit{Basic code}}:
The general code for fitting the \muhaz{} model and obtaining estimates for the hazard rate is
{\small \begin{verbatim}
	# Fit the muhaz model.  Set the maximum time so that the hazard rate is estimated
	# over all failure times, not only over observed failure times.
	muhaz.fit = muhaz(dataset$time, dataset$delta, max.time = max(dataset$time))
	
	# Obtain the estimate of the hazard rate
	muhazrate = muhaz.fit$haz.est	
\end{verbatim}}
\end{itemize}
\hrule 

\subsection{{\tt survPresmooth} package}\label{sec:presmooth}
\begin{itemize}
	\item \underline{\textit{Package and function information}}: The {\tt survPremooth} package contains functions that compute ``presmoothed" estimators of the hazard rate, cumulative hazard, and survival curve for right-censored survival data.  Estimation is implemented using the \presmooth{} function.  The presmoothing method replaces the the `0' in the censoring indicator with a non-parametric estimate through the Nadaraya-Watson regression estimator (\cite{nadaraya, watson}).\footnote{Specifications for the pre smoothing of the censoring indicator, such as the number of bootstrap resamples needed, the grid for numerical integration, and how missing values should be handled can be done through the \makefxn{control.presmooth} function.}  The kernel function for the the hazard rate is the Tanner-Wong estimator (\cite{tannerwong}).\footnote{In the \presmooth{} function, the type of kernel options are ``biweight" (default) or ``triweight" for the estimate of the hazard rate.}  The function also allows the user to correct for possible boundary effects at the beginning and/or end of the study.\footnote{This is done through the ``bound" option of the \presmooth{} command.  The default is ``none".}  The theoretical methods behind this software can be found in \cite{presmooth1,presmooth2,presmooth3,presmooth4,presmooth5,muhaz1}.
	\item \underline{\textit{Reliability}}:  May perform better with the aggregate survival and cumulative hazard measures, but did not perform well for hazard rate estimation.\footnote{All results shown in Section \ref{sec:PHsimuls} are from the ``boostrap" bandwidth selection method.  We attempted to improve upon the estimator by selecting other kernels and bandwidths, but these changes did not result in estimators that were more accurate.}
	\item \underline{\textit{Level of difficulty}}:  Moderate.
	\item \underline{\textit{Best use}}: In cases where the data set is heavily censored, presmoothing the censoring indicator can provide more information, possibly allowing for a more robust examination of the failure patterns.  While our own simulations and data analyses did not show this to be the case, there may be other circumstances where this tool improves the estimate, as it gives mass to both censored and uncensored observations.  In addition, while beyond the scope of this work, presmoothing may improve on non-parametric aggregate measures; since the Kaplan-Meier curve \cite{KMest} and the Nelson-Aalen cumulative hazard \cite{NofNA, AofNA} only have jumps where there are uncensored observations.  In the case of heavy censoring presmoothing methods may more accurately describe the survival curve or cumulative hazard. 
	\item \underline{\textit{Functions available for {\tt survPresmooth} fitted objects}}: 
	\begin{itemize}
		\item \presmooth{}: Used to fit the survival model.
		\item \makefxn{print}: The standard \R{} function for printing the fitted model.
		\item \makefxn{control.presmooth}: Returns a list of values that are used to control the parameters and output of the presmoothing of the censoring indicator (with results from the function used as a parameter in the \presmooth{} function).  
	\end{itemize}
	\item \underline{\textit{Methodological package limitations}}: 
		\begin{itemize}
			\item \textbf{Cannot accommodate covariates.}
			\item \textbf{No uncertainty bounds provided for the hazard rate.} 
			\item \textbf{Standard potential issues with kernel smoothing estimation.} See Footnote \ref{footnote:smoothprobs}, although \presmooth{} does have an option for correcting boundary issues present in smoothing.\footnote{The \presmooth{} function does attempt to correct boundary issues if the user sets the ``bound" option to ``left", ``right", or ``both", depending on which end of the study correction is needed.  Default is ``none".}
		\end{itemize}
	\item \underline{\textit{Package implementation limitations}}: 
		\begin{itemize}
			\item \textbf{No plotting or summarizing functions.} No standard \R{} functions or otherwise provided for analyzing the results of the fitted model.
			\item \textbf{Non-standard model formulation.} The standard \R{} \makefxn{Surv} and \makefxn{formula} functions re not used for input, although this not required as no covariates are included in the model.			
		\end{itemize}
\hrule
\item \underline{\textit{Basic code}}:
The general code for fitting the \presmooth{} model (no covariates) and obtaining estimates for the hazard rate is
{\small \begin{verbatim}
# Fit the pre-smoothed model.  
# Use the 'bootstrap' method for bandwidth selection.
presmooth.fit = presmooth(dataset$time, dataset$delta, estimand = 'h', bw.selec = 'bootstrap')

# Obtain the estimated hazard rate from the fitted object
presmoothhazrate = presmooth.est$estimate
\end{verbatim}}
\end{itemize}
\hrule

\subsection{{\tt timereg} package}\label{sec:timereginfo}
\begin{itemize}
	\item\underline{\textit{Package and function information}}: This package contains many functions that implement the Cox model extensions found in \cite{timereg}.  The function we examine in this manuscript is \timecox{}, which fits the proportional hazards model, but allows for covariates to be included under the non-proportional hazards assumption as well, using an extension of the traditional Cox model.\footnote{The \timecox{} function can also accommodate time dependent variables and counting process data (multiple events per subject), although that is outside the scope of this manuscript and so we do not examine this feature.} The model we use is of the form:
\begin{eqnarray}\lambda(t) = Y(t)\lambda_0(t)\exp\{X^T\beta(t)+Z^T\gamma\},\label{eqn:timecoxHazrate_generic}\end{eqnarray}
where $(X, Z)$ is a $(p+q)$-dimensional covariate, and the $\beta(t)$ are non-parametric time-varying (i.e. NPH) covariate effects, and $\gamma$ is the q-dimensional regression parameter (i.e. PH covariate effects).  In the \timecox{} function, the baseline hazard rate $\lambda_0(t)$ is modeled as $\exp\{\alpha_0(t)\}$.  Parameters are estimated using score equations.  The $\alpha_0(t)$ and $\beta(t)$ estimates are reported as cumulative estimates over the study period.\footnote{The default study period in the \timecox{} model is the first and last observed (i.e. uncensored) failure, but this can be changed by the user through the ``start.time" and ``max.time" options.}  For example, the time-varying effects of a parameter $\beta(t)$ are calculated and reported as:
\begin{eqnarray} B(t) = \int_0^t \beta(s) ds.\label{eqn:timecoxCumul}\end{eqnarray}
The formulation of this model allows for easy testing of the significance of parameters under the proportional and non-proportional hazards assumption.  The de-cumulated estimates (i.e. $\hat{\beta}(t)$) can be calculated by applying the \makefxn{CsmoothB} function to the cumulative estimates (see code below). The user may enter specifications for the estimation of the covariance matrix and on the bandwidth and degree of the local linear smoother.
	\item \underline{\textit{Reliability}}:  Moderate for de-cumulated estimates.  
	\item \underline{\textit{Level of difficulty}}:  Easy for obtaining cumulative estimates, moderate to expert to obtain de-cumulated estimates (due to lack of documentation). 
	\item \underline{\textit{Best use}}:  Because of its intuitive nature, useful examples, and fast run-time, the \timecox{} function is highly recommended for instances where testing the significance of the NPH covariate effects are important, particularly if the estimate of the hazard rate is not imperative.  The {\tt timereg} package is well-designed, user-friendly, and has plenty of associated documentation, and thus it is a very valuable piece of software for implementing a variety of complex survival models.  However, for survival modeling when the estimate of the hazard rate is needed, other more reliable functions from other packages are available for including covariates under both the PH and the NPH setting (see Discussion). 	
	\item\underline{\textit{Functions available for {\tt timereg} fitted objects}}:\footnote{In addition to \timecox{}, the package also includes a comprehensive set of complex functions, including those for an additive survival model, a semiparametric proportional odds model, fast cumulative residuals, excess risk models, and two-stage frailty models, among others.}
	\begin{itemize}
		\item \timecox{}: Used to fit the survival model.  
		\item \makefxn{plot}, \makefxn{print}, \makefxn{summary}: Standard \R{} functions used to asses the fitted \timecox{} model.
		\item \makefxn{CsmoothB}, \makefxn{Csmooth2B}: De-cumulates the cumulative estimates for $\alpha_0(t)$ and $\beta(t)$.\footnote{The \makefxn{CsmoothB} produces smoothed de-cumulated estimates of the hazard rate and the NPH covariate effects, allowing the user to control the degree of smoothing (through the parameter ``b") and the time-points at which the smoothed estimate should be calculated (specified through ``predicted.timepoints").  The function smooths the cumulative estimates of the hazard rate and NPH covariate effects, and new smoothed values are then predicted at the time points specified by the user. The difference between each sequential pair of predicted smoothed values is calculated and divided by the bin width to get the de-cumulated estimates, which should be exponentiated in the instance of the hazard rate.}
	\end{itemize}
	\item \underline{\textit{Methodological package limitations}}: 
		\begin{itemize}
			\item \textbf{Cumulative estimated covariate effects and cumulative log-hazard rate reported.} The method for reporting the hazard rate (as the cumulative log of the hazard rate) and the effects of NPH covariates (as the cumulative effect of the covariate over time) are non-standard, and therefore the interpretation of these covariates must be handled carefully.  
			\item \textbf{Necessary smoothing to obtain standard estimates.} To convert the reported hazard rate and effects of NPH covariates to the more typical de-cumulated estimates, the \makefxn{CsmoothB}, \makefxn{Csmooth2B}, or other user-defined smoothing routines must be used.  In doing this, the issues inherent to smoothing (see Footnote \ref{footnote:smoothprobs}) become part of the de-cumulated \timecox{} parameter estimates, which can dramatically impact the resulting estimate of the hazard rate or effects of NPH covariates.\footnote{The de-cumulated estimate produced by \makefxn{CsmoothB} can be dramatically affected by the degree of the smoother entered into the function.  In the results reported in the simulations, we tested a range of smoothing parameters and reported the results for the ones that produced the smallest integrated bias and integrated RMSE (see Section \ref{sec:simuls}).  In the absence of any guidance from the package manual, the smoothing parameters were visually chosen for the analyses of the ``cancer" and ``TRACE" data sets (see Section \ref{sec:ex}) as a compromise between capturing major the details in the hazard rate while maintaining a smooth estimator.}
			\item \textbf{Approximation required for 95\% confidence intervals for the de-cumulated parameters.} Use of a smoothing function for de-cumulation also requires conversion of the upper and lower 95\% confidence interval bounds provided for the cumulative estimates.\footnote{In this manuscript, we approximated the 95\% confidence interval bounds for $\exp\{\alpha_0\}$ and $\beta(t)$ by calculating the 95\% upper and lower confidence bounds for the cumulative estimates, then applying the \makefxn{CsmoothB} function to these functions. \label{footnote:timeregvar}}
		\end{itemize}
	\item \underline{\textit{Package implementation limitations}}: 
		\begin{itemize}
			\item \textbf{Undocumented de-cumulation functions.} The smoothing functions \makefxn{CsmoothB} and \makefxn{Csmooth2B} (necessary for obtaining the more standard estimated hazard rate and NPH covariate effects) are not documented in the package anywhere, and were only discovered in communicating with the author of the {\tt timereg} package.
		\end{itemize}
\hrule
\item \underline{\textit{Basic code}}: Below is the general code for fitting a \timecox{} model and obtaining the estimated cumulative hazard rate parameter $\alpha_0(t)$ (the log of the baseline hazard rate), the effects of both PH and NPH covariates, and the associated variances. 
{\small \begin{verbatim}
# Fit the timecox model for PH and NPH covariates.  
# The const() function in the formula is used to denote the PH covariates.
timecox.fit = timecox(Surv(time, delta) ~ const(covariatesPH)+covariatesNPH, data = dataset)

# Obtain estimates and variances for the PH covariates
timecox.covarPH = timecox.fit$gamma
timecox.covarPH.var = timecox.fit$var.gamma

# Obtain estimates and variances for the cumulative NPH covariates from the `cum' structure.
# The first column contains the time points at which the parameters are estimated,
# the second column contains the cumulative estimate for alpha0,
# and the remaining columns contain the cumulative estimate for the beta(t).
timecox.alpha0 = timecox.fit$cum[,2]
timecox.covarNPH.cumul = timecox.fit$cum[,-(1:2)]
timecox.covarNPH.cumul.var = timecox.fit$var.cum

# Use the CsmoothB() function to convert the cumulative estimates
# to the within bin estimates.  "predicted.timepts" are timepoints
# over the course of the study that are used to calculate the de-cumulated
# parameter estimates.  The last parameter in the function 
# (equal to '1' in this example) is a smoothing parameter.
decumulated.ests = CsmoothB(timecox.fit$cum, predicted.timepts, b = 1)
timecox.hazardrate = exp(decumulated.ests[,2])
timecox.covarNPH = decumulated.ests[,-(1:2)]
\end{verbatim}}
\end{itemize}
\hrule
\subsection{{\tt YPmodel} package}
\begin{itemize}
	\item \underline{\textit{Package and function information}}: This package contains various inference procedures associated with the \ypmodel\footnote{While \makefxn{YPmodel} is the primary function for estimation (according to the user manual), the user cannot obtain the estimates and error bounds from the fitted object produced using this function, and may only view the estimated hazard ratio on the \R{} console.  In order to obtain and use the actual parameter estimates and the error bounds from the fitted object, \ypmodel{} must be used.} function, which estimates the log-hazard ratio between two non-proportional hazard groups using a semi-parametric model.\footnote{This model contains the proportional odds model as a sub-model, but this is outside the scope of this work and its performance is not investigated.}  This model is based on work found in  \cite{ypmodel1, ypmodel2}, and  is frequently referred to as a short-term and long-term hazard ratios model when referring to a treatment and control group, as the two survival curves may cross if treatment has adverse affects in the beginning of a study (performing worse than a control group), but has benefits over a longer period of time.  This method adjusts the common Cox model $\lambda_T(t) = \theta\lambda_C(t)$, where `T' refers to the treatment group, `C' refers to the control group, and $\lambda_C(t)$ is the baseline hazard, and is written as:
	$$\lambda_T(t) = \frac{\theta_1\theta_2}{\theta_1+(\theta_2-\theta_1)S_C(t)}\lambda_C(t); \theta_1, \theta_2 > 0.$$
	With this parameterization, $\theta_1$ and $\theta_2$ can be interpreted as the short-term and long-term hazard ratios.  Parameters are estimated through a pseudo maximum likelihood function\footnote{The number of iterations for estimation can be specified by the user, as well as initialization and boundary values for the ratio estimates.} which is then connected to martingale theory for variance estimation.  
	\item \underline{\textit{Reliability}}:  Not reliable due to various issues in package design.
	\item \underline{\textit{Best use}}: This method is designed for the case of crossing hazards, and may be ideal in some instances other than those examined in this manuscript.  The methodology behind this function is easy for most \R{} users to grasp as it is closely related to the Cox proportional hazards model, the parameters are estimated through the pseudo-maximum likelihood, and model testing is available.  
	\item \underline{\textit{Level of difficulty}}: Expert.
	\item \underline{\textit{Functions available for {\tt YPmodel} fitted objects}}:  
	\begin{itemize}
		\item \makefxn{YPmodel}: The main function for fitting the survival model.  Results and estimates can only be displayed on the console and cannot be not returned as objects to the user.  Internal error messages appear when the different components of the fitted model are accessed.
		\item \ypmodel: Performs the same operation as \makefxn{YPmodel}, but values can be returned as objects to the user.
		\item \makefxn{YPmodel.lackfittest}: Performs lack of fit tests on \makefxn{YPmodel}.
		\item \makefxn{YPmodel.adlgrk}: Calculates the p-value of the adaptively weighted log-rank test.
		\item \makefxn{plot}, \makefxn{print}, \makefxn{summary}:  The standard  \R{} functions that are technically available for the fitted object, however, they do not work as they should (see \underline{\textit{Package implementation limitations}}).
	\end{itemize}
	\item \underline{\textit{Methodological package limitations}}: 
		\begin{itemize}
			\item \textbf{No inclusion of PH covariates.} Only one NPH covariate is allowed in this model.
			\item \textbf{No hazard rate estimate(s) produced.} The log-ratio of the two hazards is the only estimate that can be produced using this method.
		\end{itemize}
	\item \underline{\textit{Package implementation limitations}}:
		\begin{itemize}
			\item \textbf{Standard \R{} functions do not work properly.}  While the standard \makefxn{plot}, \makefxn{print} and \makefxn{summary} functions are nominally available for the package, all three functions produce the same four plots and summaries.  In addition, use of the functions produces an internal error message:
	{\small \begin{verbatim}
	> plot(YPmodel.nph)
	Error in 1:n : argument of length 0
	\end{verbatim}}
			\item \textbf{Two functions perform the same estimation routine.} Both \makefxn{YPmodel} and \ypmodel{} perform the same estimation routine, but only the second one provides output the user can access.\footnote{The main function for the package (\makefxn{YPmodel}) does not allow the user to use the results in the fitted object, and internal error messages are displayed when they are accessed.  To obtain estimates or error bounds of the log-ratio or other fitted model parameters, the user must implement \ypmodel{}. The reason for two different functions is not clear.}
			\item \textbf{User must name the columns of the data set to be analyzed `V1', `V2', `V3'.} Any other set of names given to the data set will cause the routine to crash, so each data set must be re-named before using with this function.
			\item \textbf{Standard \R{} \makefxn{Surv} and \makefxn{formula} functions are not used for input}, although this is not an issue as no covariates are included in the model.
			\item \textbf{Difficult to find the extra documentation.} In the downloaded \R{} package, an additional documentation file (analogous to a vignette) is available, and provides additional useful information on how to use each function in the package.  However, this additional file is not loaded in the typical manner so is difficult to find.\footnote{The file is not loaded into the package as a vignette (i.e. typing \texttt{vignette("YPmodel")} into the \R{} console does not display the file).  Instead, the file must be accessed from the {\tt YPmodel} package folder downloaded from \cranr.  There is a note in the ``description" section of the package documentation that this file exists, but will likely not be noticed by a majority of \R{} users.}
		\end{itemize}
\hrule
\item \underline{\textit{General code}}: 
Below is the general code for fitting the \ypmodel{} and obtaining the estimated log-ratio and the confidence bounds.  
{\small \begin{verbatim}
# Column names for the dataset must be 'V1', 'V2', and 'V3', and columns
# must be ordered as: time, censoring variable, non-proportional covariate
names(dataset) = c('V1', 'V2', 'V3')
# Fit the model
YPmodel.fitNPH = YPmodel.IntervalBands(dataset)

# Obtain the hazard ratio estimate
YPmodel.covarNPH = YPmodel.fitNPH$hr
# Obtain the upper and lower bounds of the 95% point-wise confidence intervals
YPmodel.covarCI = cbind(YPmodel.fitNPH$low3, YPmodel.fitNPH$upp3)
# Obtain the upper and lower bounds of the 95% simultaneous confidence bands
YPmodel.covarBand = cbind(YPmodel.fitNPH$low22, YPmodel.fitNPH$upp22)
\end{verbatim}}
\end{itemize}
\hrule

\section{Simulation Results} \label{sec:simuls}

As one of the main goals of this work is to compare the accuracy and reliability of the different functions, we first examined the estimates produced by each of the nine functions discussed in Section \ref{sec:summary} in a simulated data setting.  In the simulations, survival data was generated with covariates included under the proportional hazards (PH) and the non-proportional hazards (NPH) assumptions. Under each of the two scenarios (PH and NPH), we generated 100 data sets, each with 1000 people stratified equally among the different covariate groups, and failure times were generated over the course of ten years.  Functions that could not accommodate covariates\footnote{\bgp{}, \muhaz{}, and \presmooth{} in the PH setting, and \ypmodel{} in the NPH setting} were examined using the baseline subjects only.

The accuracy of the estimates produced by each function were assessed through visual examination, as well as through bias and the square root of the mean square error (RMSE) calculations. For each package, the bias of the hazard rate (obtained from the 100 fitted models) was calculated at each time point $t_j$ as 
\begin{eqnarray} bias(\hat{h}(t_j)) = 1/100 \sum_{i=1}^{100} (h(t_j)-\hat{h}(t_{ij})),\label{eqn:bias} \end{eqnarray}
and the RMSE of the hazard rate was calculated at each time point as 
\begin{eqnarray} RMSE(\hat{h}(t_j)) = \sqrt{1/100 \sum_{i=1}^{100} (h(t_j)-\hat{h}(t_{ij}))^2}.\label{eqn:rmse}\end{eqnarray}  For ease of calculation and comparison of the bias and RMSE values, the time grid over which the hazard rate was estimated was set to 32 equally spaced bins (i.e. 3.75 months per bin) for functions that allowed this specification.\footnote{\baysurv{}, \bshaz{}, \lddp, \emrh{}, \muhaz{}, and \presmooth{}, as well as \timecox{} through the predicted time points entered into \makefxn{CsmoothB}.}  This number was chosen as a compromise between fine-grain analysis of hazard rate patterns while remaining computationally efficient.

Functions evaluated under the PH setting were \baysurv, \bgp, \bshaz, \emrh, \lddp, \muhaz, \presmooth, and \timecox, and functions evaluated under the NPH setting were \baysurv, \emrh,  \lddp,\timecox, and \ypmodel.

\subsection{Hazard rate estimation with a proportional hazards covariate}\label{sec:PHsimuls}
We first assessed the hazard rate estimates produced by all functions (except \ypmodel{}\footnote{\ypmodel{} can only examine the effects of one NPH covariate and does not produce an estimate of the hazard rate.}) and the estimated effects of one binary covariate included under the proportional hazards assumption.  The failure times in the PH simulations were generated from a true hazard rate that had a sharp initial peak, indicating a high failure rate in the beginning of the study period, followed by a period of steady reduced risk. The shape of the true hazard rate was designed to assess which estimators were able to capture sharp peaks near the beginning of the study period as well as steady decline across a period with fewer observations. A treatment covariate was included in the model under the proportional hazards assumption, with a true $\beta$ equal to -0.5.  The average rate of censoring was 63\%.  Among functions that did not provide any estimate of the covariate effects,\footnote{\bgp{}, \muhaz{}, \presmooth{}} the hazard rate for the baseline group was estimated using only the 500 baseline subjects.  

\begin{figure}
	\centering
	\includegraphics[width=6.5in]{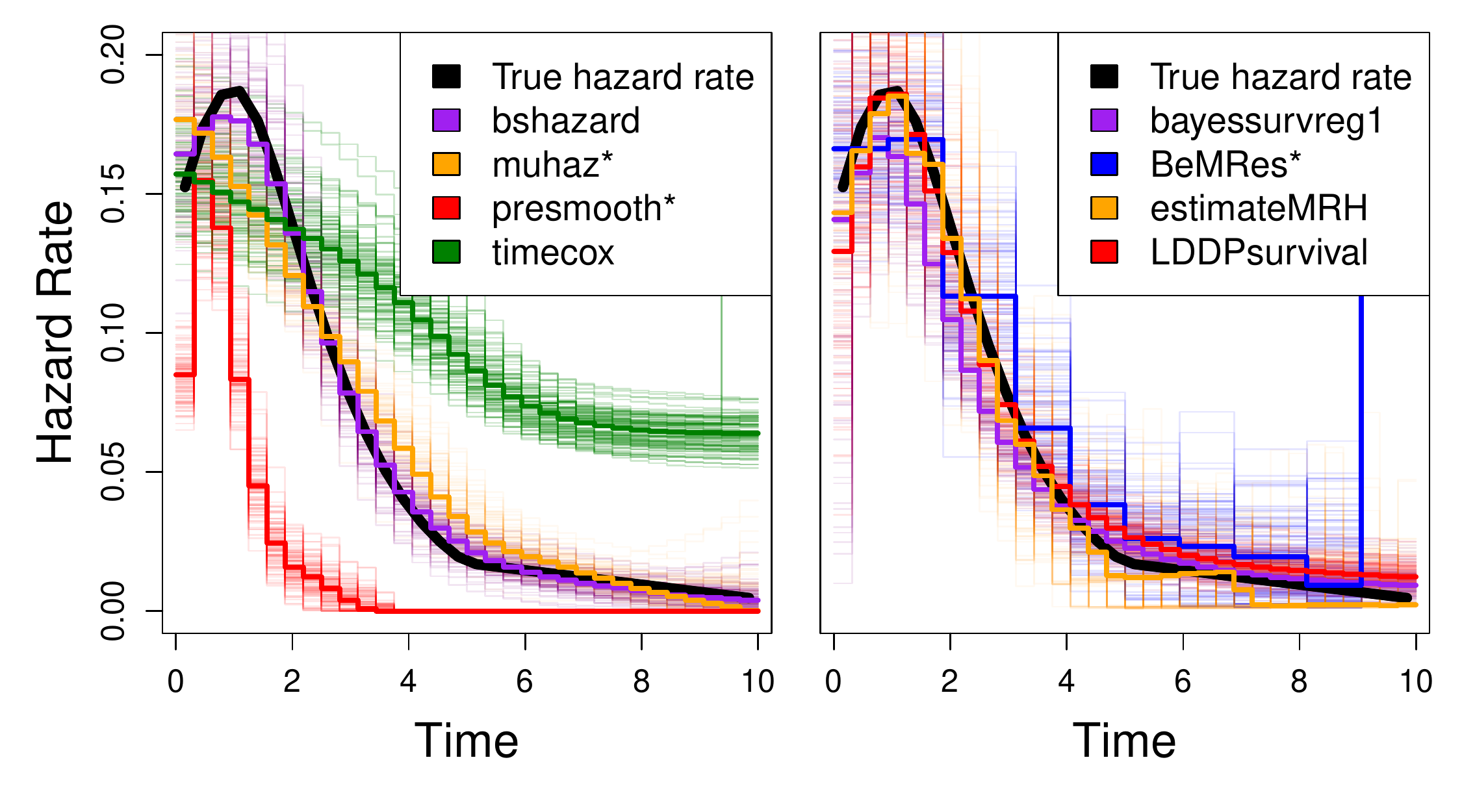}
	\includegraphics[width=6.5in]{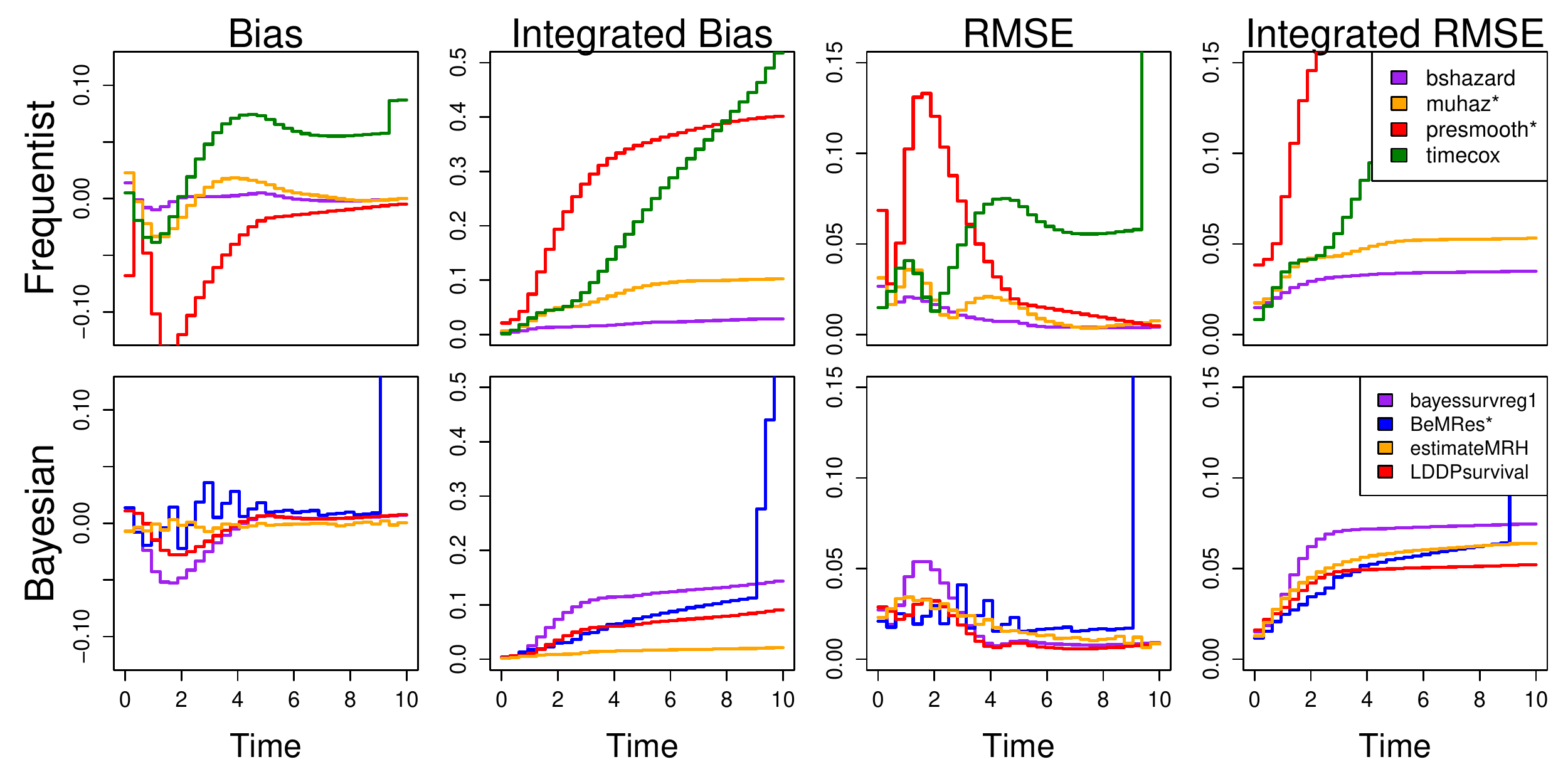}
	\caption{\footnotesize  Comparison of the estimated baseline hazard rate for the PH simulations.  For each method, the top graphs show the estimated hazard rate for each of the 100 simulations (the ``cloud" of lines) with the mean of the 100 simulations superimposed with a darker line, and contrasted against the true hazard rate, which is shown in black.  The top left figure shows the results of the functions based on frequentist methods, where it can be observed that \bshaz{} does the best job in capturing the entire curve.  On the top right, among the functions based on Bayesian methods, it can be observed that \emrh{} and \lddp{} produce the best results.  The bottom graphs show the bias, RMSE, integrated absolute bias, and integrated RMSE calculations for each function.  Among frequentist methods, \bshaz{} performs the best according to all measures.  Among the Bayesian approaches, \emrh{} and \lddp{} perform the best, with smaller bias values for \emrh{} and smaller RMSE values for \lddp. A `*' denotes results that are based on baseline subjects only.}	
\label{fig:compareHazEsts_PHsimul}
\end{figure}

\begin{figure}
	\centering
	\includegraphics[width=6.75in]{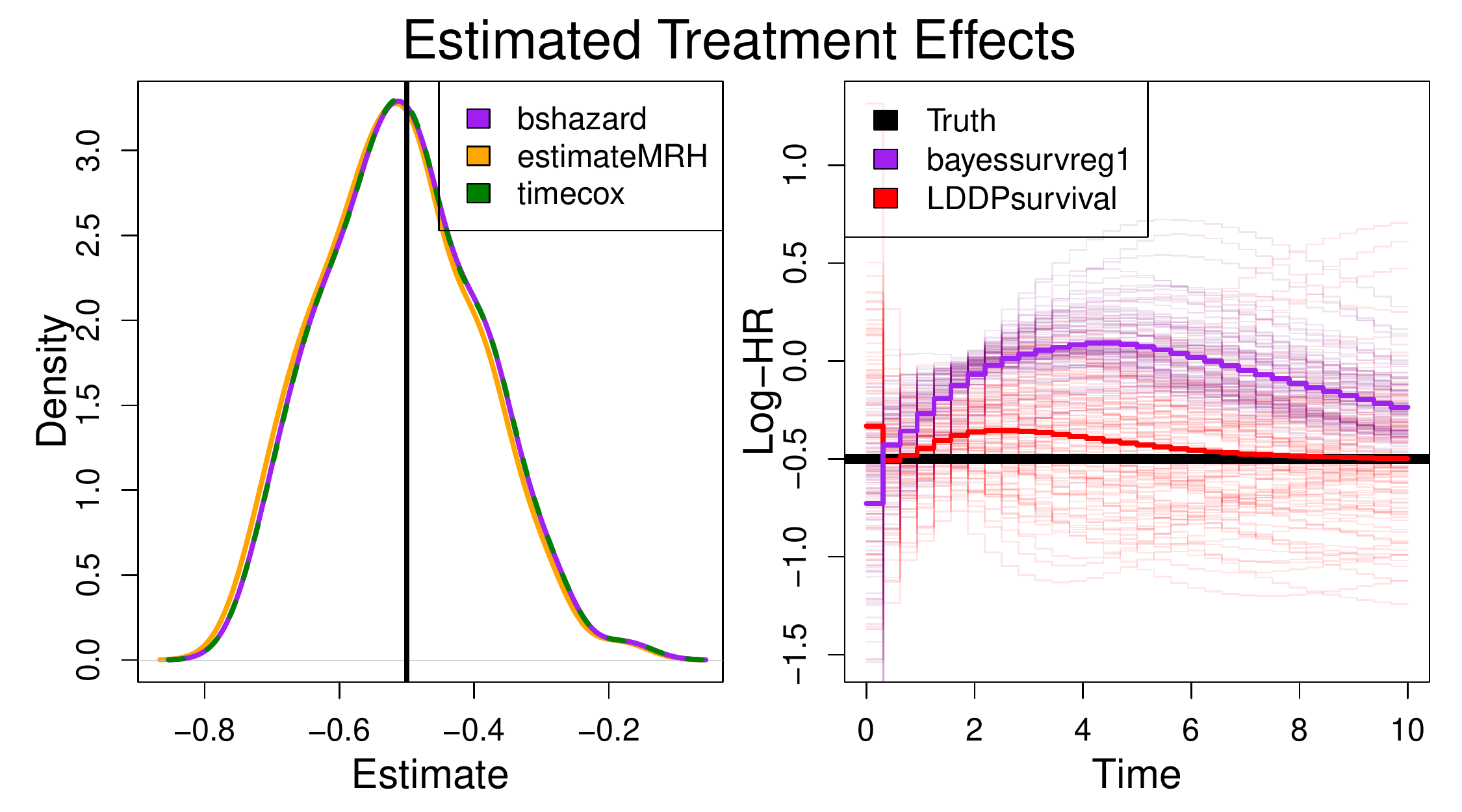}
	\caption{\footnotesize  Comparison of the estimated covariate effects for the PH simulations.  The left graphs show densities  (calculated using the \makefxn{density} function in \R{}) of the 100 treatment covariate estimates produced by \bshaz{}, \emrh{}, and \timecox{}, with the true value (equal to -0.5) denoted by the solid vertical line.  The estimate densities appear identical between the three methods.  The right graph shows the estimated time-varying log-hazard ratio between the treatment hazard rate and the baseline hazard rate for the PH simulations, contrasting the \baysurv{} and \lddp{} models, both of which have different covariate interpretations than the standard Cox PH interpretation.  The results of each of the 100 simulations are shown in lighter colors, and the mean of the 100 estimates at each time point are shown with solid colors.  The true value of $\beta$ is shown in black.  We can see that the results of the \lddp{} model are close to proportional, but the results of the \baysurv{} model deviate quite a bit, with a large hump in the middle of the study.  A `*' denotes results that are based on baseline subjects only.}	
\label{fig:compareBetaEsts_PHsimul}
\end{figure}

The results of the simulations can be observed in Figures \ref{fig:compareHazEsts_PHsimul} and \ref{fig:compareBetaEsts_PHsimul}.  The top graphs of Figure \ref{fig:compareHazEsts_PHsimul}  display the estimated hazard rate for each of the 100 simulated data sets (the ``cloud" of lines) as well as the mean of the 100 estimates (the darker line), with the true hazard rate superimposed in black.  The bottom graphs of Figure \ref{fig:compareHazEsts_PHsimul} show the bias, integrated absolute bias, RMSE, and integrated RMSE for all methods.  Among the frequentist methods, \bshaz{} provides estimates closest to the true hazard rate, and also has the smallest bias, integrated absolute bias, RMSE and integrated RMSE. While \muhaz{} has trouble capturing the initial peak in the hazard rate, it performs well in the last two-thirds of the study period.  The results of the \presmooth{} and \timecox{} functions are not as accurate, and both have departures dramatically different than the true curve.  Among the Bayesian models, it is observed that the \emrh{} and \lddp{} models perform the best, with the MRH method showing a smaller bias and the LDDP method showing a smaller RMSE.  While \bgp{} performs well throughout most of the study, the jump towards the end (which is likely due to the small sample size towards the end of the study period) makes its use questionable in studies with periods of sparsely observed failures.  The \baysurv{} method performs well, although the true initial peak in the hazard rate is not fully captured.  On average, it appears that the Bayesian models tend to have lower bias and RMSE values than the frequentist models, with the exception of the accurate results produced by \bshaz{} and perhaps \muhaz{}.  

Results of the estimated covariate effects (Figure \ref{fig:compareBetaEsts_PHsimul}) show little difference among the methods that produce an estimated PH covariate effect (right graph).  Both the \baysurv{} and \lddp{} models have different interpretation of covariate effects due to the way the models are parameterized, with hazard log-ratios that change over time for all covariates included in the model, and the results for these models are shown in the left graph.  The estimated effect of the treatment covariate was calculated by taking the log of the ratio of the treatment hazard rate and the baseline hazard rate,\footnote{The hazard rate for each covariate group can be done using the \makefxn{predict} function in the {\tt bayesSurv} package, and by specifying ``zpred" in the \lddp{} function}  producing an estimated treatment covariate effect that changes over time.  Because the treatment covariate effect was simulated under the proportional hazards assumption, we would expect the log-ratio of the two hazards to be constant throughout the study. This seems to be the case for the \lddp{} model, which shows a constant hazard log-ratio that is very close to the true value of the proportional hazards covariate used to generate the data.  However, the results of the \baysurv{} model deviate from the truth, with an estimate that tends to have a hump in the middle of the study and that does not seem to show two proportional hazard rates, which  may be due to smaller sample sizes after the initial peak in failures.


\subsection{Hazard rate estimation with non-proportional hazards covariates}

In addition to PH simulations, we also evaluated the estimates produced by the five functions that accommodate a non-proportional covariate effect.\footnote{\baysurv, \emrh{}, \lddp{}, \timecox{}, and \ypmodel}  In the NPH simulations, we included one binary covariate (treatment) under the non-proportional hazards assumption, and one binary covariate (gender) under the proportional hazards assumption.  The shape of the hazard rate for treatment group 1 was the same as in the PH simulations.  The shape of the hazard rate for treatment group 2 was flat, allowing subjects in this group to have a lower initial hazard rate when compared to treatment group 1, but a higher hazard rate as the study progressed, inducing crossing hazards.  The true gender covariate effects value was $\beta = -0.5,$ and for the purposes of discussion we assume that males were the baseline subject group.  The average percent censored in treatment group 1 was 63\%, and the average percent censored in treatment group 2 was 30\%.    Because the \ypmodel{} function does not accommodate proportional hazards covariates, NPH log-ratio estimates were calculated for the 500 subjects in the baseline PH group (i.e. males) only.  In addition, this method does not provide an estimate of the hazard rate, so this comparison is not made for this function.

\begin{figure}
	\centering
	\includegraphics[width=6.75in]{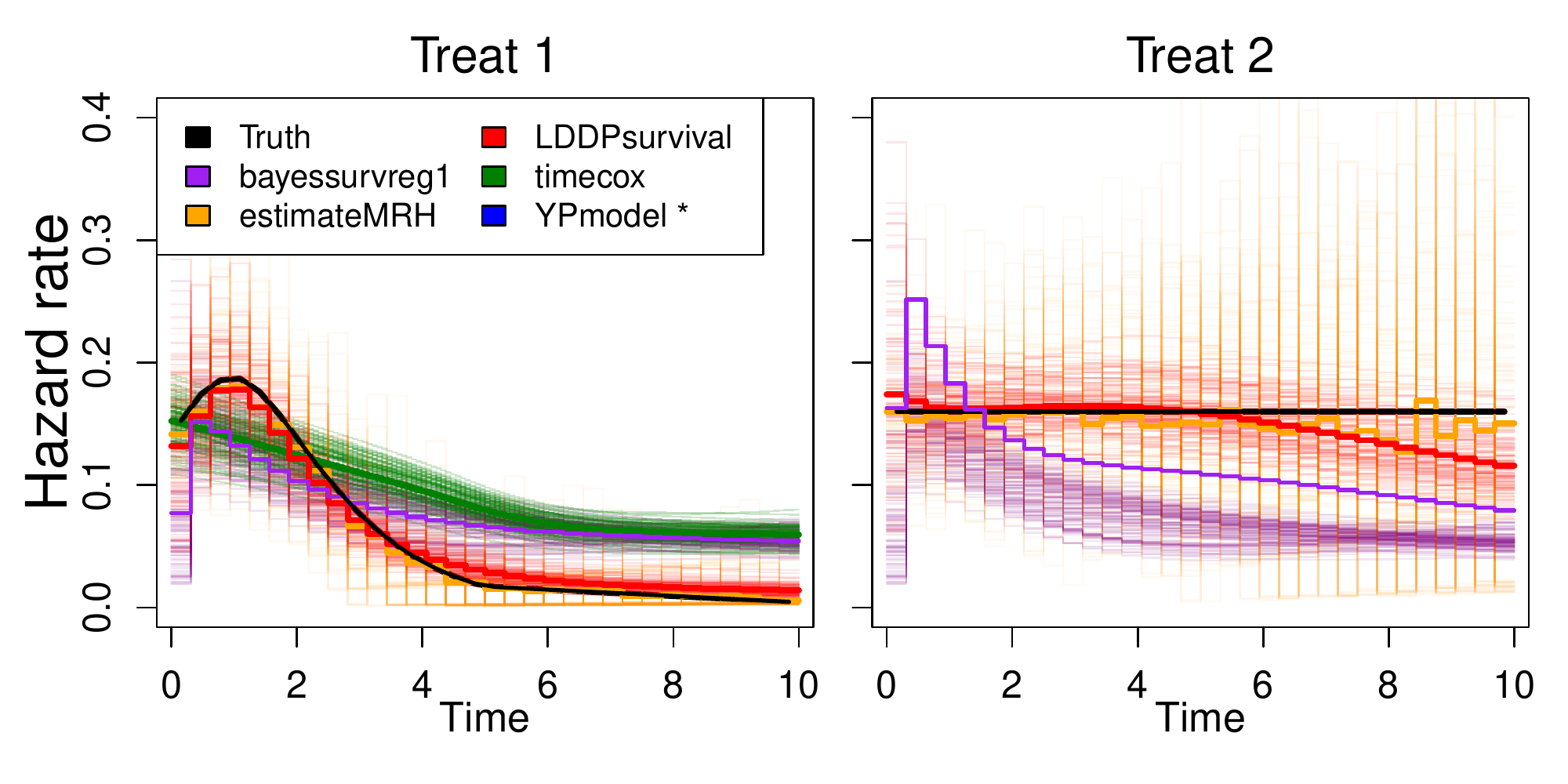}
	\caption{\footnotesize  Comparison of the estimated hazard rate for the NPH simulated data, contrasting models that can accommodate at least one covariate under the non-proportional hazards assumption. Results are from 100 simulated data sets where the true hazard rate for treatment group one has an initial sharp peak and then a declining rate afterward, and the second treatment group has a flat hazard rate over the course of the study, causing the hazard rates to cross.  For each method, estimates from each of the simulations are shown with a ``cloud" of lighter lines, and the mean of the 100 simulations is superimposed with a darker line. The true value the data was generated from is shown in black. The left graph shows the estimated hazard rates for each function (with the exception of \ypmodel, as it does not provide an estimate of the hazard rate) for treatment group 1 and the right group shows the estimated hazard rates for treatment group 2. The \emrh{} and \lddp{} functions seem to do the best job capturing the true shape of both hazard rates, although the \lddp{} estimate for the hazard rate of treatment group 2 is too small towards the end of the study, when the number of observed failures is small.  The \baysurv{} model struggles to capture the shapes, although this is expected as AFT models cannot accommodate crossing hazards. The results of \timecox{} and \ypmodel{} both show estimates that are smoother than the true values.  A `*' denotes results that are based on baseline subjects (males) only.}	
\label{fig:compareHazEsts_NPHsimul}
\end{figure}
\begin{figure}
	\centering
	\includegraphics[width=5in]{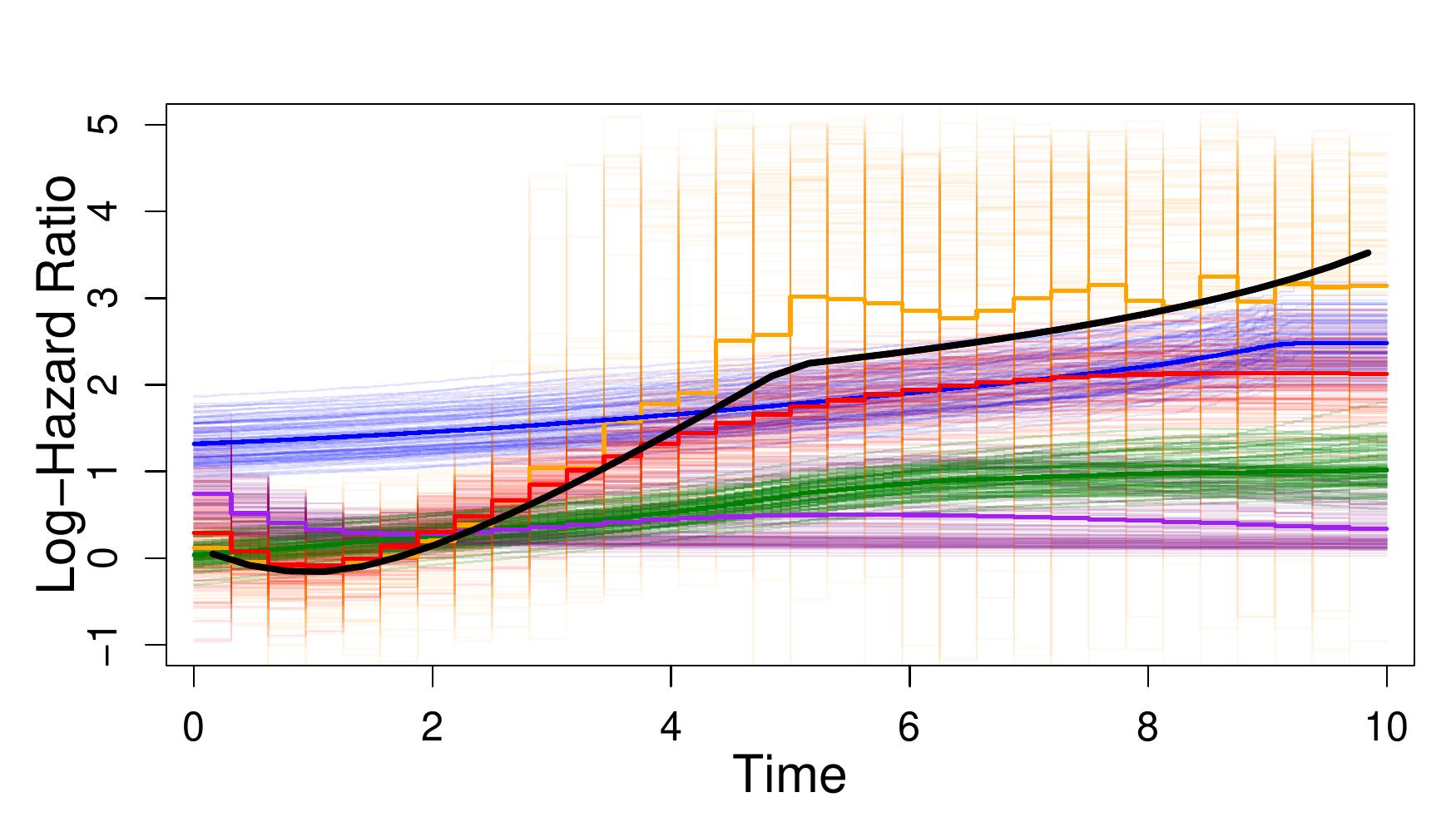}
	\includegraphics[width=5in]{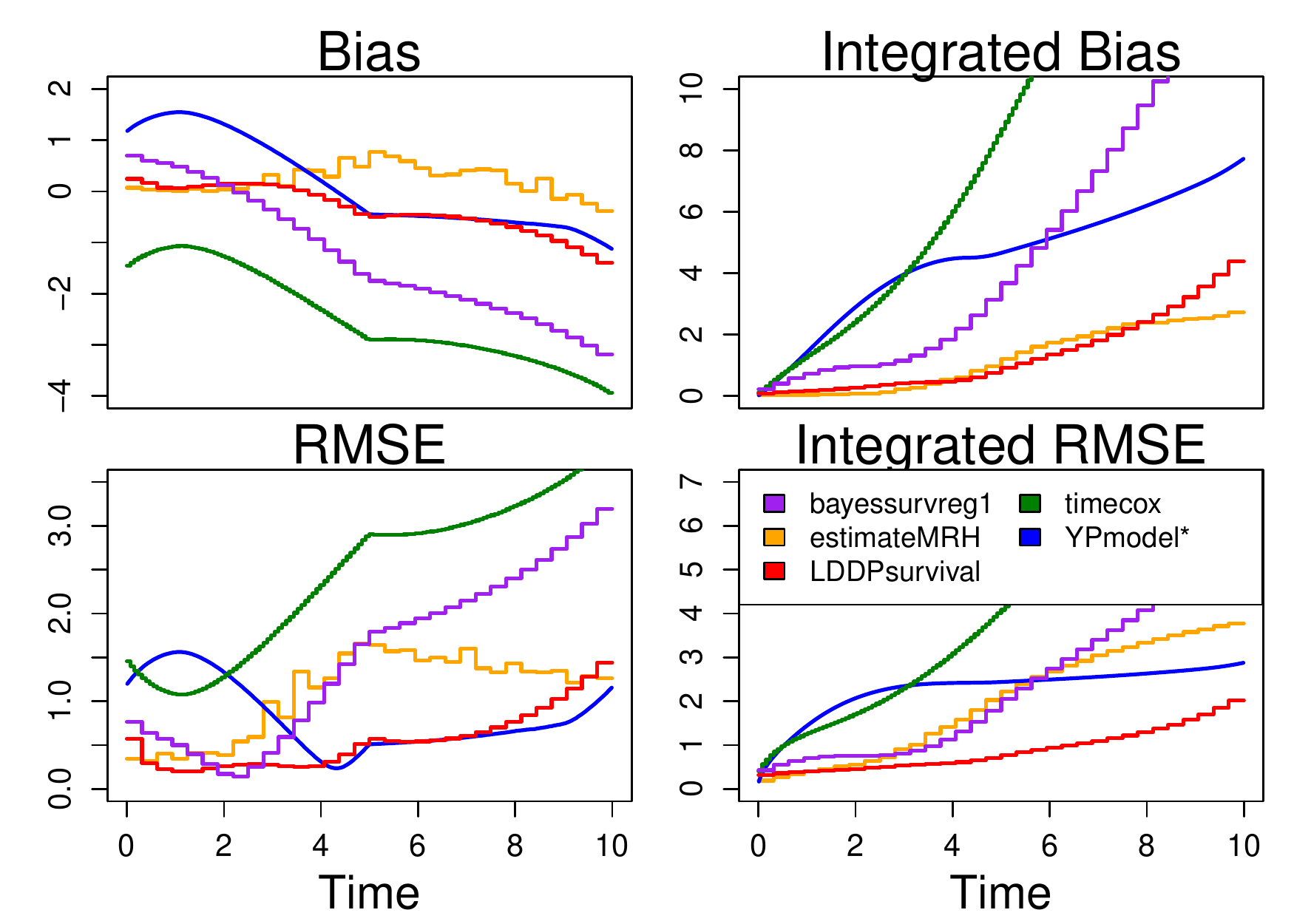}
	\caption{\footnotesize Comparison of the estimated NPH treatment covariate effect for the NPH simulated data (top) and the bias, integrated absolute bias, RMSE, and integrated RMSE (bottom graphs). In the top figure, for each method, treatment log-ratio estimates from each of the simulations are shown with a ``cloud" of lighter lines, and the mean of the 100 simulations is superimposed with a darker line. The true value the of the treatment log-ratio is shown in black.  The \emrh{} and \lddp{} functions seem to best capture the true shape of the log-hazard ratio of the treatment effect, while the other three models are flatter and do not capture the curve.  The bias and RMSE comparison (bottom graphs) of the estimated treatment covariate effect for 100 NPH simulated data sets show that as with the PH simulations, across the bias and integrated absolute bias calculations, the \emrh{} model performs best, while across RMSE and integrated RMSE calculations, \lddp{} performs best.  A `*' denotes results that are based on baseline subjects (males) only.}	
\label{fig:compareNPHBetaEsts_NPHsimul}
\end{figure}

\begin{figure}
	\centering
	\includegraphics[width=6.75in]{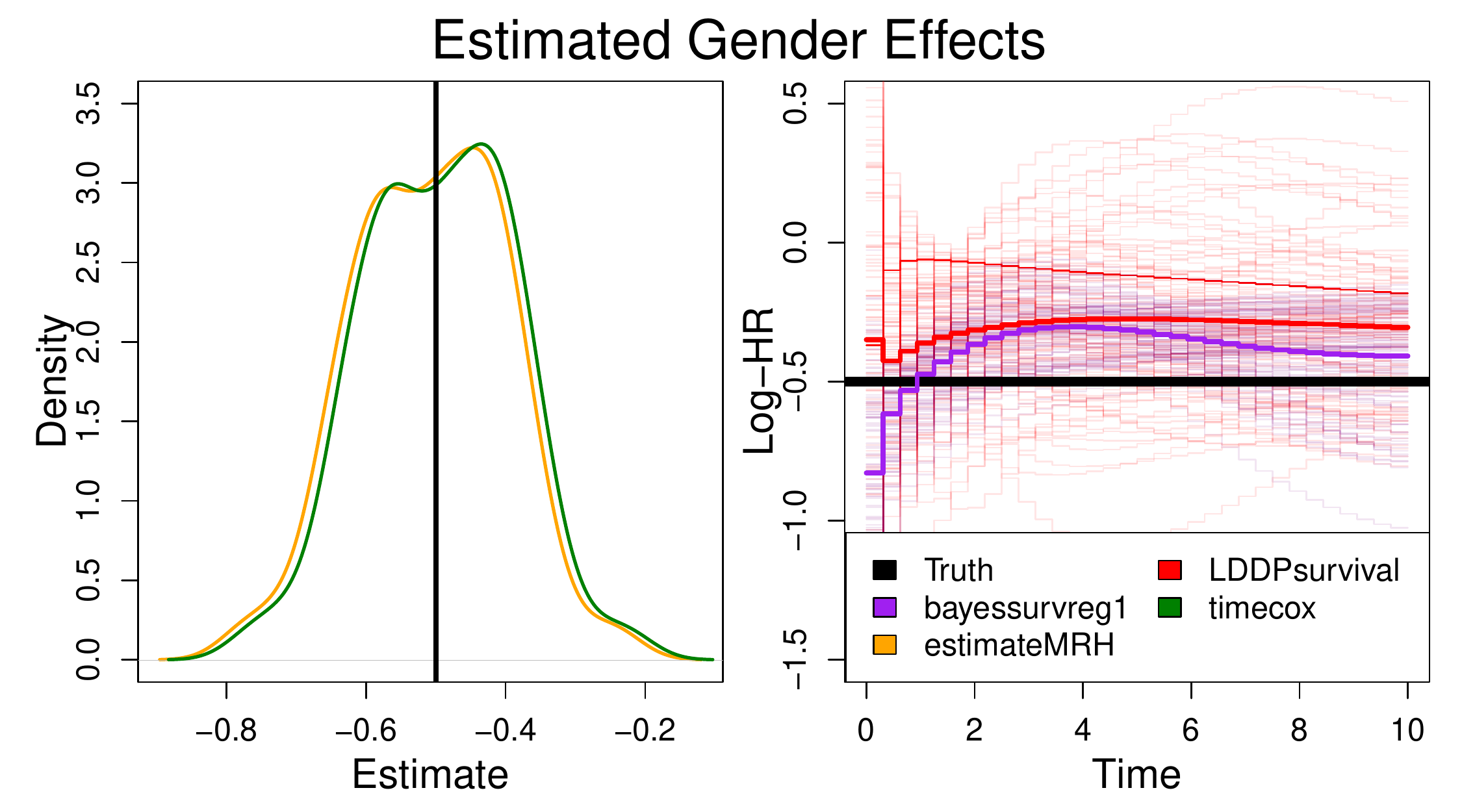}
	\caption{\footnotesize The estimated log-hazard ratio of gender effects for the 100 NPH simulations.  On the left, the density of the 100 estimates (calculated using \makefxn{density} in \R) is shown for the \emrh{} and \timecox{} models, with very little difference observed between the two methods.  On the right, the estimates for the \baysurv{} and \lddp{} models are shown separately, as the covariate interpretation for these two models is different than the traditional Cox model, and the log of the ratio of the hazard for males and the hazard for females is allowed to change over time. The results of each of the 100 simulations are shown in lighter colors, and the mean of the 100 estimates at each time point are shown with solid colors.  We can see that both models estimate a fairly proportional relationship between the two hazards, although on average the ratio is larger than it should be for the \lddp{} model (around -0.35 instead of -0.5).  In addition, the \baysurv{} results show a hump in the first half of the study, with an estimate that is too small in the first year.  The results of the \lddp{} model are on average more stable throughout the study. }	
\label{fig:comparePHBetaEsts_NPHsimul}
\end{figure}

The results of the NPH simulations can be observed in Figures \ref{fig:compareHazEsts_NPHsimul}, \ref{fig:compareNPHBetaEsts_NPHsimul}, and \ref{fig:comparePHBetaEsts_NPHsimul}.  Figure \ref{fig:compareHazEsts_NPHsimul} shows the hazard rate estimates for treatment 1 and treatment 2 for the different models.  We see that the \emrh{} and \lddp{} functions seem to produce the most accurate estimates of the true hazard rates (and therefore treatment effect), although the \emrh{} model does a better job estimating the hazard rate for treatment group 2 towards the end of the study.  The, \timecox{} model does not capture the initial peak and an estimated hazard rate too high when compared to the true hazard rate.  While the \baysurv{} model does a better job capturing the initial peak of the treatment 1 hazard rate, it also falsely captures an initial peak for the treatment group 2 hazard rate, which should be flat.  The comparison of the estimated log-hazard ratio for these methods can be seen in Figure \ref{fig:compareNPHBetaEsts_NPHsimul} (top).  The log-hazard ratio for the NPH covariate highlights the same results shown in the previous figure, and also contains the results for \ypmodel{}, which show flat estimates that do not capture the initial dip in the first two years.  The bottom graphs of the figure show the bias, integrated absolute bias, RMSE, and integrated RMSE for the different models. As with the PH simulations, the \emrh{} model results have the lowest bias, but the \lddp{} model has the lowest RMSE.  It is also likely the \baysurv{} model, which is an AFT model, struggles as it is not designed to accommodate crossing hazard rates.

The estimated gender effects for the 100 simulations are shown in Figure \ref{fig:compareHazEsts_NPHsimul}, which shows the densities of the 100 estimates for the \emrh{} and \timecox{} models on the left, and the estimates for the \baysurv{} and \lddp{} models on the right.  The densities show very little difference between the \emrh{} and \timecox{} models.  Estimated covariate effects for the \baysurv{} and \lddp{} functions are interpreted differently, and are shown in the right graph of the figure. The estimated effect of gender was calculated by taking the log of the ratio of the hazard rate for females and the hazard rate for males.  As with the PH simulations, we would expect this log-ratio to be a constant value at -0.5.  However, while both models seem to show a somewhat proportional relationship between the two hazards (with stronger linear patterns in the \lddp{} model), there are a few issues. As with the proportional hazards simulations, the \baysurv{} results show a hump, with initial estimates being too small and deviating from the truth quite a bit.  The \lddp{} results are more stable and have a more linear shape, but the overall covariate effect is larger than the true effect (closer to -0.35 as opposed to the true -0.5).


\section{Real data example}\label{sec:ex}
In addition to assessing the repeated performance of the functions through simulations, we also tested the functions' performance in a real data analysis setting.  This gave us the opportunity to investigate additional exploratory features of each package, and to compare the error bounds of the parameter estimates provided by the packages (if available).  We analyzed two data sets from actual studies (originally found in \R{} packages), and performed one analysis with covariates included under the proportional hazards assumption, and another analysis with covariates included under both the proportional hazards and non-proportional hazards assumption.  We limited our real data analyses to functions that incorporate covariates.

In the PH analysis, we examined the ``cancer" data set (available in the \texttt{survival} package), and in the NPH analysis, we examined the ``TRACE" dataset (available in the \texttt{timereg} package).   For simplicity, we categorized all continuous variables in both analyses.  For both examples, the initial Cox proportional hazards models (fitted using \makefxn{coxph}) along with the \makefxn{stepAIC}\footnote{While stepwise model selection methods are simplistic and are not meant to be used in state-of-the-art biological model building, they provide adequately selected models for our comparison purposes.} function were used to select covariates to be included in the models.  

\subsection{PH example} \label{sec:PHexample}

The ``cancer" data set from the \texttt{survival} package was analyzed in order to compare the exploratory features and results when all covariates were included under the proportional hazards assumption.  This data set contains 228 patients with advanced lung cancer from the North Central Cancer Treatment Group \cite{cancerdataset}, and includes measurements of the survival time in days, as well as other demographic and biological information for each patient.  Based on the Cox proportional hazards models and step-wise AIC methods, the selected predictors included  gender, weight loss in the last six months, and physician rated ECOG (Eastern Cooperative Oncology Group) performance scores \cite{ecog}.  Weight loss was categorized by quartiles, and ECOG scores were grouped into categories with subjects rated as either 0, 1, or 2/3, with 0 representing the best possible performance, and 2/3 representing a poor score.  The data set is 28\% censored, with a median observed failure time of 256 days (range is 5 to 1022 days).  The baseline group ($n = 16$) were males with ECOG scores equal to 1 and a weight loss measure in the first quartile (weight loss $\le$ 0 kg).  In this analysis we compared results from \baysurv, \bshaz, \emrh, \lddp, and \timecox.

\subsubsection{Parameter estimate comparison}

\begin{figure}
	\centering
	\includegraphics[width=6in]{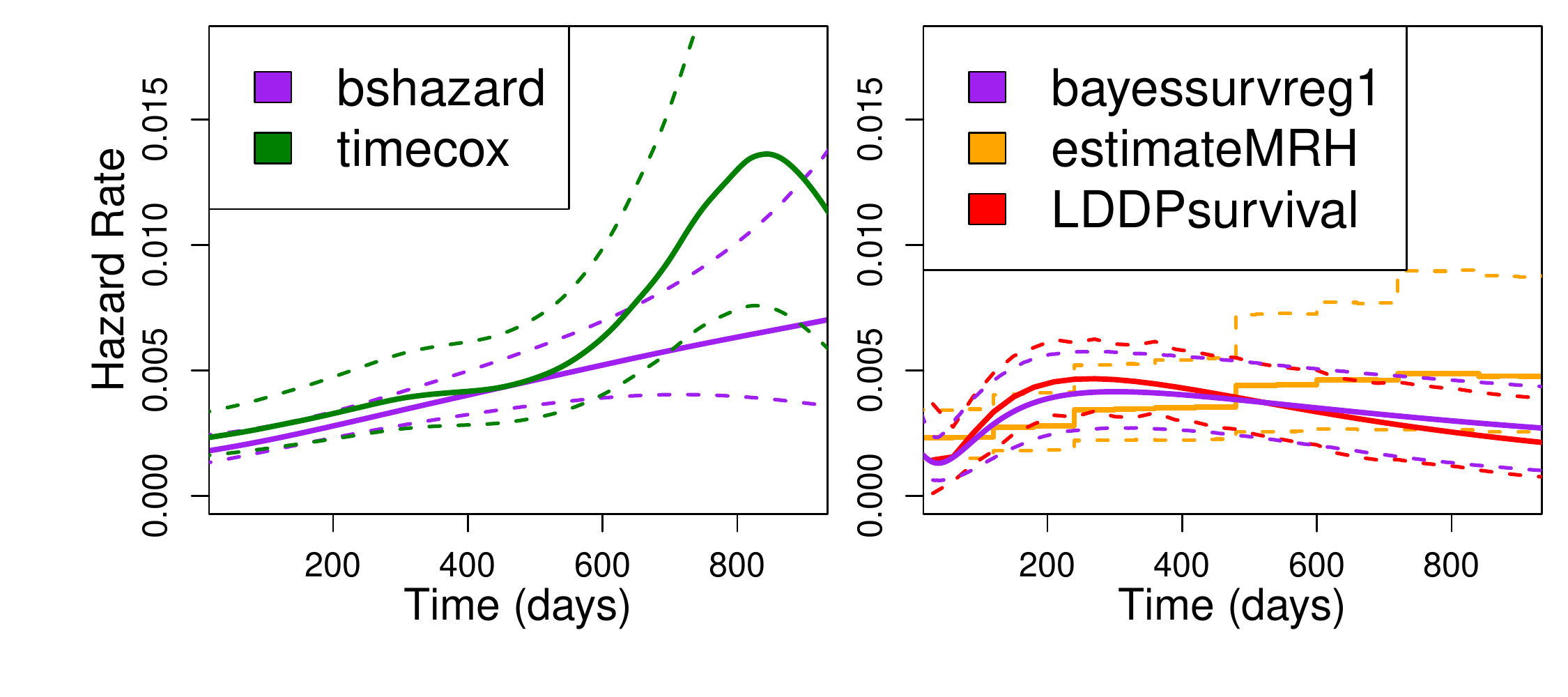}
	\includegraphics[width=4in]{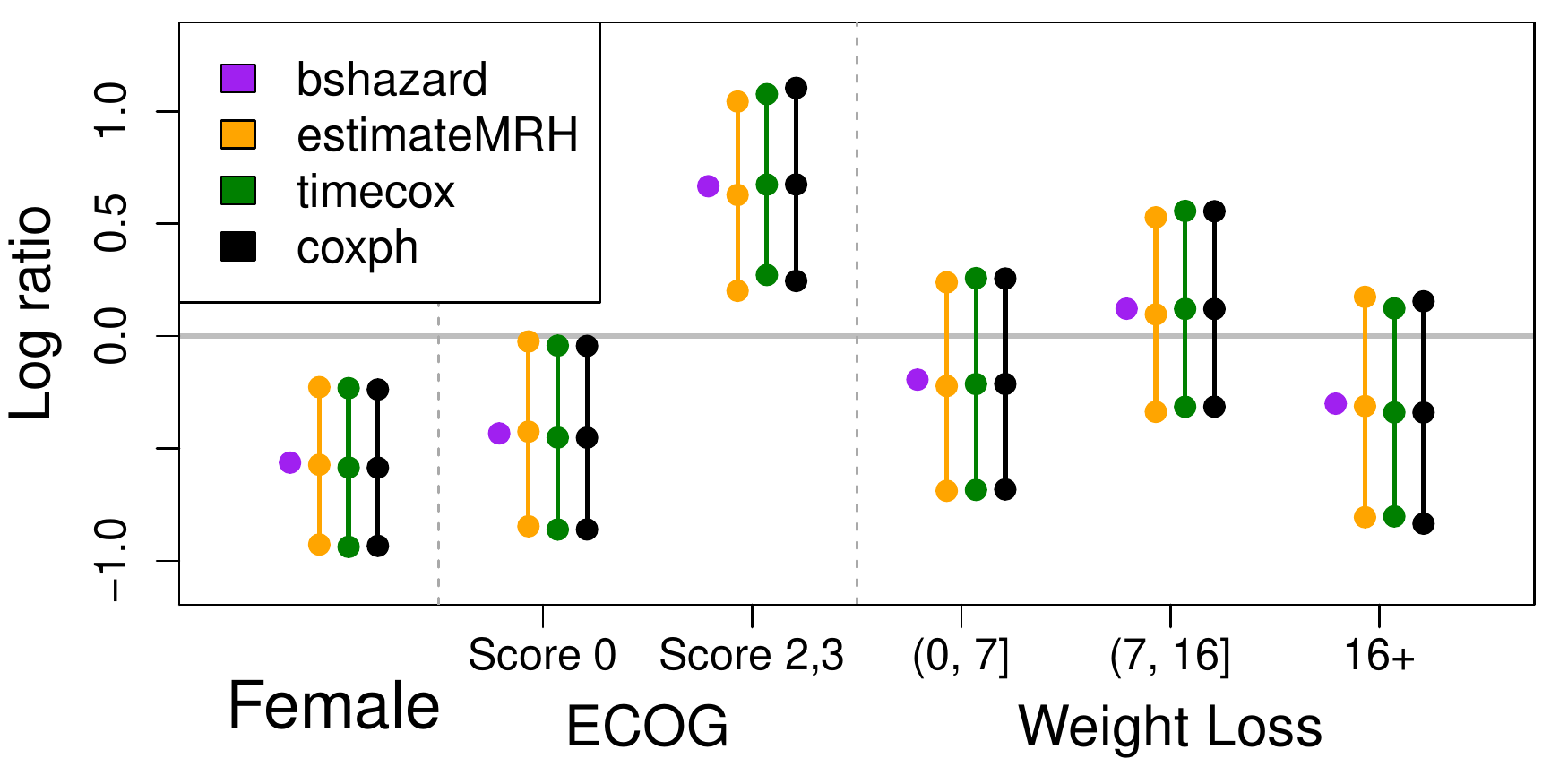}
	\caption{\footnotesize Comparison of the estimated baseline hazard rates and covariate effects among the different functions examined in the PH example using the ``cancer" data in the \texttt{survival} package.  Estimated covariate effects are shown with thick solid lines, and 95\% interval bounds are shown with dashed lines (credible intervals for \baysurv, \emrh, \lddp, and confidence intervals for \bshaz{} and \timecox).  The top two figures show the estimated hazard rate for the frequentist models (left) and the Bayesian models (right).  The \baysurv{} and \lddp{} models show similar hazard rate estimates and their interval bounds, with a slight initial hazard peak at around 300 days followed by a gradual decrease. The other functions show a flat to mildly increasing hazard rate, with varying heights achieved towards the end of the study based on the model.  The estimated PH covariate effects are shown in the bottom graph, contrasted against the estimates from the standard Cox model (implemented via the \makefxn{coxph} function in the {\tt survival} package).  All PH estimates and bounds look similar, although bounds for the \bshaz{} model are not shown since they are not available from the fitted object.  The results from the \baysurv{} and \lddp{} models are not shown as they are not interpreted in the same manner as the other models (see Figure \ref{fig:Ests_PHex_bayessurv}) and Footnote \ref{footnote:DP_prederror}.}	
\label{fig:Ests_PHex}
\end{figure}

\begin{figure}
	\centering
	\includegraphics[width=6in]{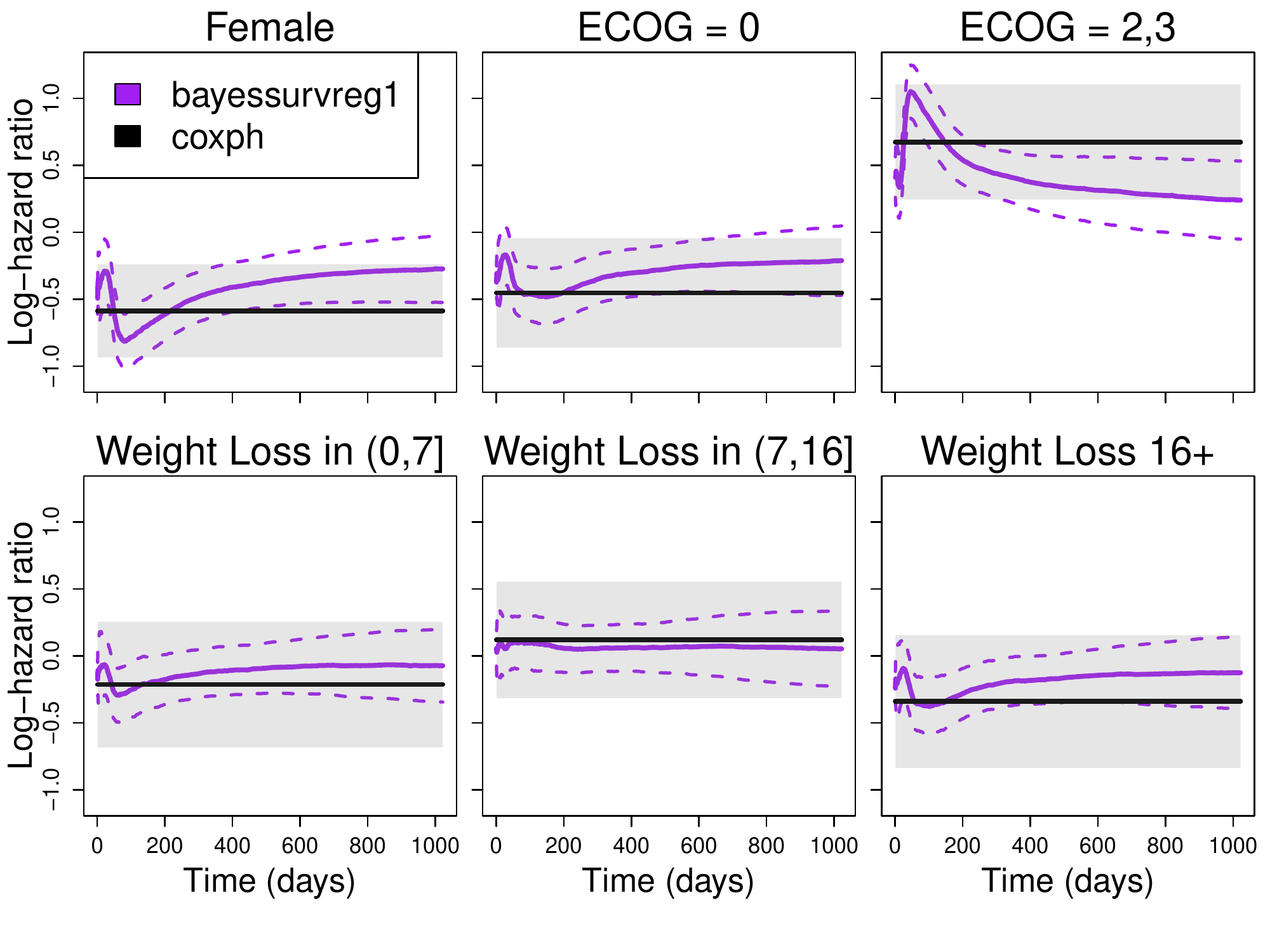}
	\caption{\footnotesize The estimated log-hazard ratio of the hazard rate for each covariate and the baseline hazard rate for the \baysurv{} model, examining the ``cancer" data set (used for the PH example).  Because of the parameterization of AFT models, the interpretation of the effect of a covariate is different than the standard Cox model, and the log-ratio between the hazard rates for two covariate groups is time-varying over the course of the study.  These figures show the estimated log-hazard ratio between the hazard rate calculated for the covariate group (such as ``females") compared to the baseline group, with the estimate denoted with a solid line, and approximated 95\% confidence intervals shown with dashed lines (see Footnote \ref{footnote:boundsapproxBS}).  The \baysurv{} estimates are contrasted with the standard Cox model estimates (solid black line) and 95\% confidence intervals (grey box), calculated using \makefxn{coxph} in the {\tt survival} package.  Some estimated covariate effects are quite similar to the Cox estimates, particularly those related to the amount of weight lost.  However, some estimated covariate effects (such as the ECOG covariates)  show greater departures.  All \baysurv{} estimated covariate effects remain in the 95\% confidence intervals produced by the Cox model.}	
\label{fig:Ests_PHex_bayessurv}
\end{figure}

The estimates of the hazard rates and the covariate effects can be observed in Figures \ref{fig:Ests_PHex} and \ref{fig:Ests_PHex_bayessurv}.  The top two graphs of Figure \ref{fig:Ests_PHex} show the estimated hazard rate for the two frequentist approaches on the left, and the three Bayesian approaches on the right.\footnote{The frequentist approaches include \bshaz{} and \timecox{}, and the Bayesian approaches include \baysurv{}, \emrh{}, and \lddp{}.}  The results show very similar estimates for the \baysurv{} and \lddp{} survival models, with a slight peak at approximately 300 days, followed by a gradual decline.  The remaining three models show a slightly increasing hazard rate, with the end height varying based on the method.  Both the \bshaz{} and \emrh{} models show a fairly linear increase in the hazard rate and are on relatively the same scale.  The \timecox{} model estimate, while also increasing, contains a distinguishable large peak at 800 days.  This hump, however, may be  an artifact of the smoothing required to reasonably de-cumulate the estimates.  We also observe that the bounds for the \baysurv{}, \lddp{} and \bshaz{} models seems narrowest, and the bounds for the \timecox{} model seems widest, particularly towards the end of the study.  Both the \bshaz{} and \timecox{} interval bounds are likely narrower than they should be due to the methods used to approximate the lower and upper interval bounds (see Footnotes \ref{footnote:bshazvar} and \ref{footnote:timeregvar}).  The bottom graph in in Figure \ref{fig:Ests_PHex} shows the estimated PH covariate effects, where it can be seen that there are almost no differences between the estimates and interval bounds (credible intervals for \emrh{} and confidence intervals for \timecox{} and the standard Cox model, implemented via \makefxn{coxph} in the \R{} {\tt survival} package), with bounds not included for the \bshaz{} model as none are provided by the function.\footnote{An additional note: after the \bshaz{} routine finished running, the following internal error message is reported: \\
\centerline{\begin{tt}
Error in aggregate.data.frame(x\$raw.data, by = list(x\$raw.data[, 5:(5 +  : 
 	arguments must have same length
\end{tt}}\\
although it does not appear to have any known effects on the hazard rate of covariate parameter estimates.}

 The estimated time-varying log-hazard ratio of the \baysurv{} model are shown in Figure \ref{fig:Ests_PHex_bayessurv}, contrasted against the PH (i.e. constant) estimates created from a standard Cox model.  Some covariates (particularly the weight loss covariates) show a very proportional effect over time, with slight changes only in the beginning of the study.  Other covariates (such as the ECOG covariates), show much bigger departures from the Cox estimate.  However, over the course of the study, the \baysurv{}  log-hazard ratio estimates do not fall outside the 95\% confidence intervals for the Cox model covariates.  Similar graphs can technically be calculated for the \lddp{} model, however when we attempted this we routinely received error messages\footnote{The error message we received was: {\tt Error in LDDPsurvival.default(... : 
  matrix is not pd in chol subroutine and were unable to predict hazard rates for all the covariate subgroups.\label{footnote:DP_prederror}}} and so are unable to show the log-hazard ratios for all covariate groups.

\subsection{NPH example} \label{sec:nphex}

The ``TRACE" data set, found in the \texttt{timereg} package, was analyzed to compare model results when covariates were included under both the PH and the NPH assumptions. This data set contains survival information on 1,877 patients who have had a myocardial infarction, along with various associated risk factors \cite{tracedataset}, with survival recorded in years.  Based on the preliminary Cox proportional hazards analysis and step-wise AIC methods, our chosen model included three binary variables indicating if the subject had clinical heart pump failure (CHF), diabetes, or ventricular fibrillation (VF), as well work metabolic index (WMI, a numeric measurement of the heart pumping effect), gender, and age. Age and WMI were categorized by quartiles.  The data set was 49\% censored, with a median observed failure time of 6.1 years (range is $< 0.01$ to 8.5 years).  Baseline subjects were those who had no VF or CHF, a WMI between 1 and 1.5, were 60 years or younger, and who did not have diabetes.  

The proportionality assumption of each variable was tested using \makefxn{cox.zph}, and results showed both CHF status and VF status did not satisfy the proportional hazards assumption and so were included in the models as NPH covariates.  Because two NPH variables were required for the modeling, \ypmodel{} was not included in the analysis as this function can only accommodate one non-proportional covariate.  Additionally, we were unable to successfully estimate all the parameter effects with \lddp,\footnote{In this particular example, we were unable to include both the CHF and VF covariates in the \lddp{} model without crashing the routine, even with all other covariates eliminated.  However, in the PH case we were able to include all 6 covariates, so it is difficult for us to predict in advance how many covariates the \lddp{} model will be able to accommodate.} so this function is omitted from this analysis.  Thus, in this example, we compare results from \baysurv, \emrh, and \timecox{} fitted models.

\subsubsection{Estimate Comparison}
The likelihood functions for all three models are very different.  The \baysurv{} and \timecox{} models estimate the covariate effects for CHF (regardless of VF status), and VF (regardless of CHF status).  However, the \emrh{} model estimates the covariate effects for each strata of the intersection of the two covariates: the baseline group, CHF only group, the VF only group, and the CHF+VF group (see Equation (\ref{eqn:MRHlikelihood})).  This poses a seeming disadvantage for the \emrh{} model, as the estimated NPH covariate effects rely on smaller sample sizes, and will inevitably have larger credible intervals.  Meanwhile, the \baysurv{} and \timecox{} estimates can be used to obtain the results for the CHF+VF group, but still retain the advantage of estimation using a larger sample size.

\begin{figure}
	\centering
	\includegraphics[width=6.5in]{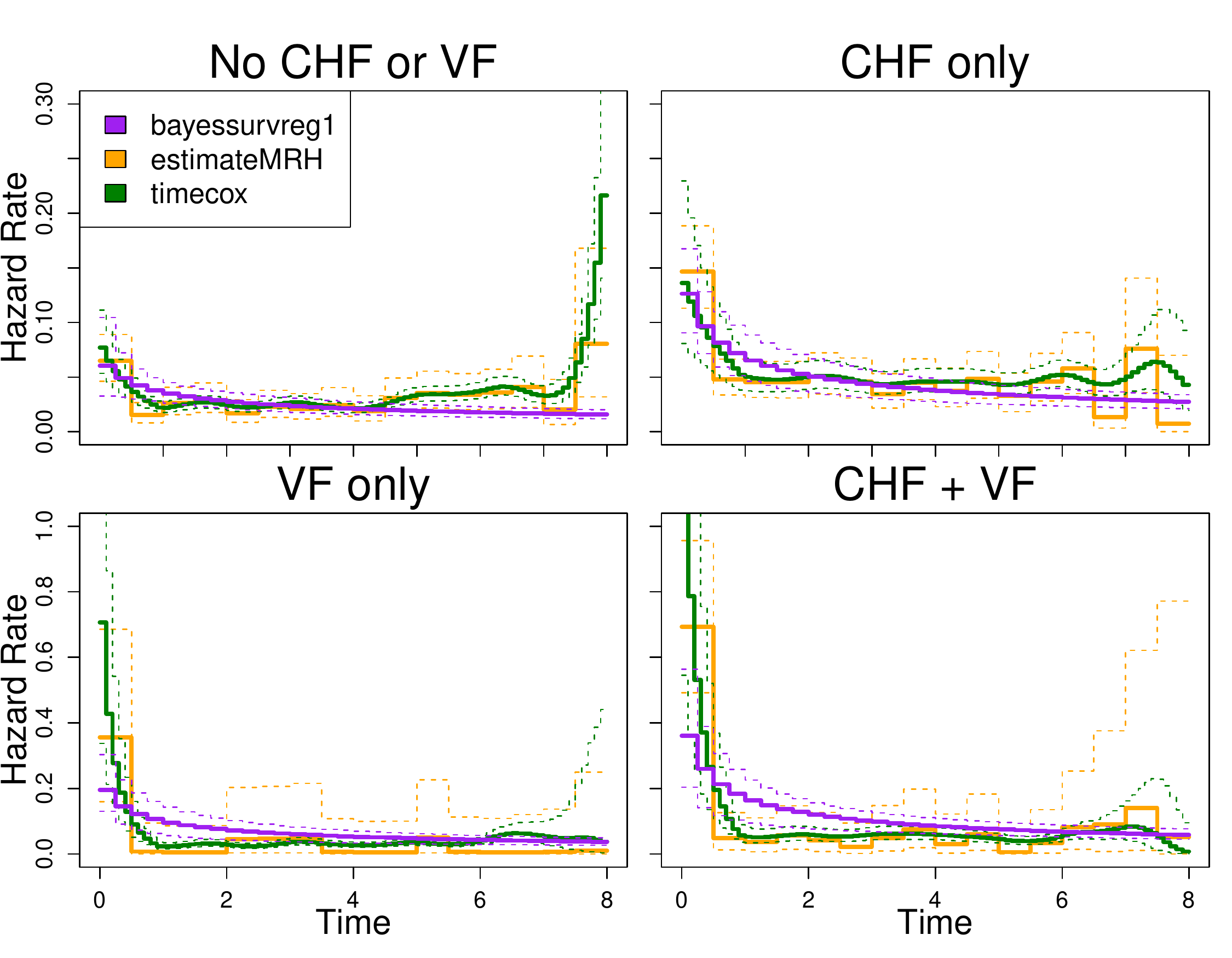}
	\caption{\footnotesize  Comparison of the estimated hazard rates of the NPH covariates in the ``TRACE" data set (i.e. the NPH example), examining the baseline group, CHF only group, VF only group, and CHF+VF group for the \baysurv, \emrh, and \timecox{} models.  Solid lines denote the hazard rate estimate, and dashed lines denote the 95\% interval bounds (credible intervals for \baysurv{} and \emrh, and confidence intervals for \timecox).  The \timecox{} bounds are also smoothed values of the upper and lower limits (see Section \ref{sec:timereginfo}, \underline{\textit{Functions available for {\tt timereg} fitted objects}}).  The \baysurv{} estimates all decrease over time.  The \emrh{} and \timecox{} models are more in agreement with each other, with increasing hazards for the baseline and CHF only groups.  The \emrh{} model has larger 95\% credible intervals when compared to the other two models, as the hazard rates for each NPH group are modeled separately.   Note that the y-axes of the upper and lower figures are different.}
\label{fig:compareNPHhazrate}
\end{figure}
\begin{figure}
	\centering
	\includegraphics[width=6.5in]{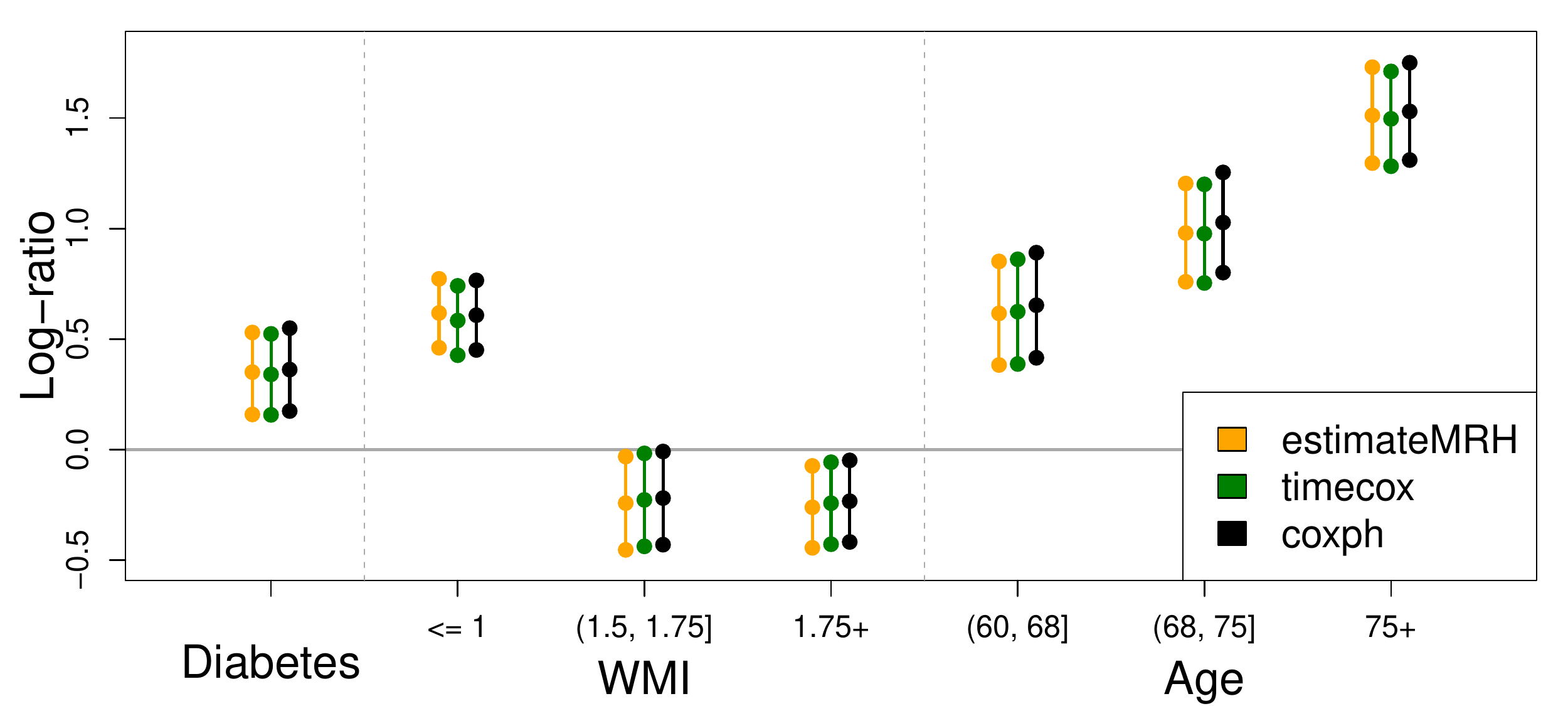}
	\caption{\footnotesize  The estimated PH covariate effects and 95\% intervals  in the ``TRACE" data set (i.e. the NPH example), comparing the results of the \emrh{}, \timecox{}, and standard Cox (calculated using \makefxn{coxph} in the {\tt survival} package) models.  Credible intervals were calculated for the \emrh{} model, and confidence intervals were calculated for the \timecox{} and \makefxn{coxph} models).  The differences between the estimates are minimal.  Results for the \baysurv{} model are not shown as the interpretation for AFT model parameters is different (see Figure \ref{fig:compareNPHbetasex_bayessurv}).}
\label{fig:compareNPHbetasex}
\end{figure}
\begin{figure}
	\centering
	\includegraphics[width=6.5in]{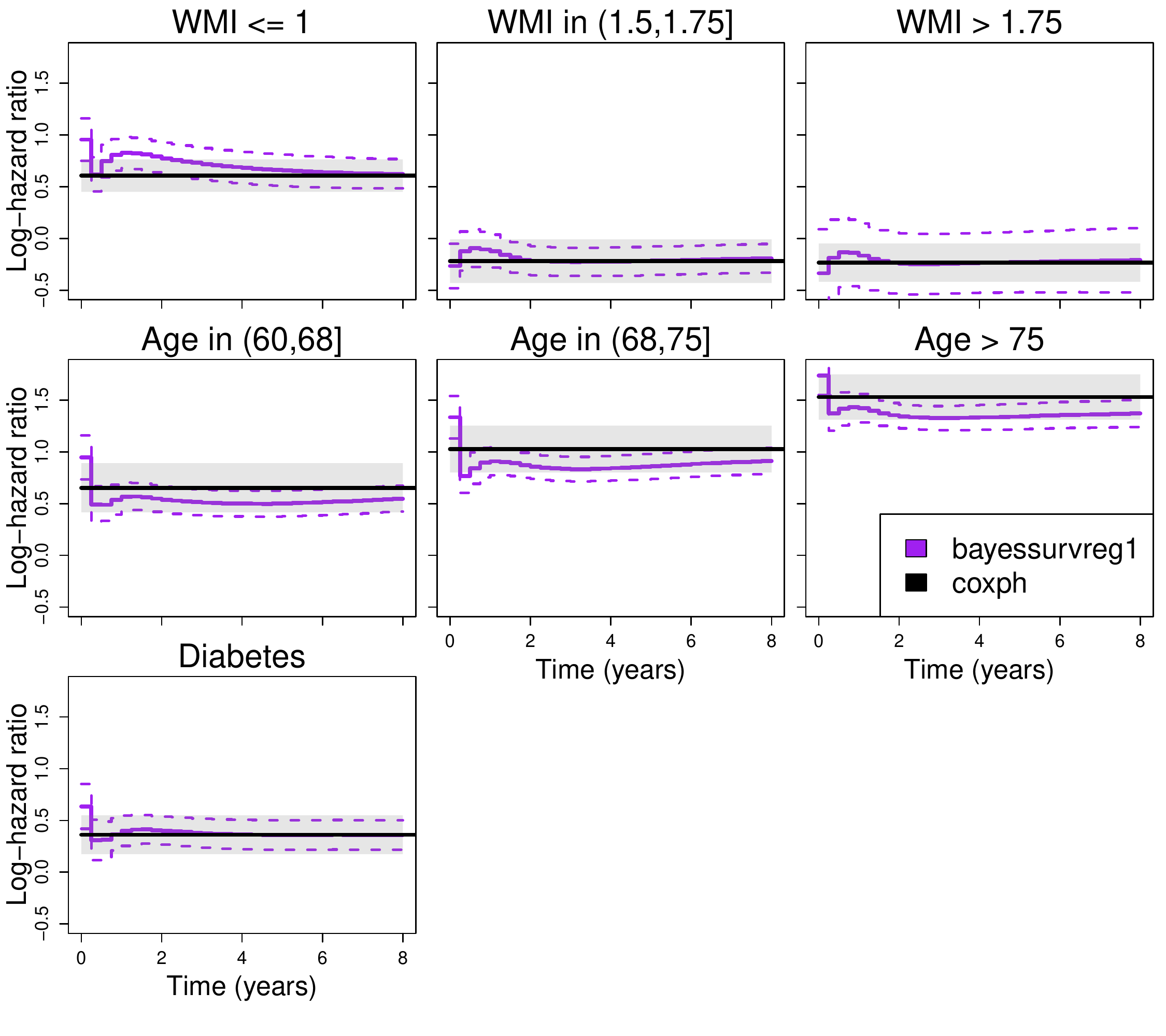}
	\caption{\footnotesize  Comparison of the \baysurv{} time-varying estimated log-hazard ratios for the different covariate groups in the ``TRACE" data set (i.e. the NPH example), with the estimates denoted with the solid purple line and the 95\% approximated confidence intervals (see Footnote \ref{footnote:boundsapproxBS}) marked with dashed purple lines.  The estimates are compared to the standard Cox model estimate (denoted with the solid black line, calculated using \makefxn{coxph} in the {\tt survival} library) and the associated 95\% confidence intervals denoted with the grey box.  Unlike the PH example in the previous section, many of the \baysurv{} log-ratio estimates go outside the Cox model confidence intervals, although a few (such as the effects of diabetes) remain quite similar and constant.}
\label{fig:compareNPHbetasex_bayessurv}
\end{figure}

Figure \ref{fig:compareNPHhazrate} shows the estimated hazard rates for all three models, as well as the 95\% error bounds (credible intervals for \emrh{} and \baysurv{}, and confidence intervals for \timecox).  In the \baysurv{} model results, we see estimates for all four hazard rates that start high initially and that continue to decrease over time, with estimates smoother than the other two models.  The \emrh{} and \timecox{} estimates  follow each other closely, with the \timecox{} model estimating higher initial and end peaks for the baseline group, the VF group, and the CHF+VF group when compared to the \emrh{} model, although this is in part due to the choice of the smoother.  The \emrh{} model has larger 95\% credible intervals when compared to the other two models, which makes sense since the NPH covariate effects are estimated using a smaller sample size.   There is little difference in the estimated effects of covariates included under the PH assumption (see Figure \ref{fig:compareNPHbetasex}).

The time-varying estimated log-hazard ratio of the covariate effects for the \baysurv{} model are shown in \ref{fig:compareNPHbetasex_bayessurv}, contrasted against the standard Cox model estimates and 95\% confidence intervals.  Unlike the PH example in the previous section, many of the \baysurv{} estimated log-hazard ratios go outside the Cox 95\% confidence intervals.  This includes all of the age effects, as well as the effects of individuals with a WMI $\le$ 1.  The effects of diabetes and the other WMI groups are similar to the PH estimates.

\subsubsection{Parameter testing}
The formulation of the \timecox{} model has strength in its ability to test the significance of the parameters, particularly in the NPH case where one might want to perform a  statistical test on whether the inclusion of a covariate under the NPH assumption is valid.  In this instance, using the \makefxn{summary} function on a \timecox{} fitted object shows inference for the covariate parameters as well as the hazard rate (which is labeled ``Intercept" in the \timecox{} output):
{\small
\begin{verbatim}
> summary(fit.timecox)
Multiplicative Hazard Model 
Test for nonparametric terms 
Test for non-significant effects 
            Supremum-test of significance p-value H_0: B(t)=0
(Intercept)                         55.00                   0
chf                                  8.43                   0
vf                                   5.55                   0

Test for time invariant effects 
                  Kolmogorov-Smirnov test p-value H_0:constant effect
(Intercept)                          1.48                       0.175
chf                                  1.78                       0.103
vf                                   1.85                       0.191
                    Cramer von Mises test p-value H_0:constant effect
(Intercept)                          6.36                       0.116
chf                                 10.60                       0.037
vf                                  16.80                       0.026

Parametric terms :     
                              Coef.    SE Robust SE     z P-val
const(diabetes)               0.340 0.093     0.120  2.83 0.005
const(as.factor(wmiCateg))b1  0.584 0.080     0.100  5.84 0.000
const(as.factor(wmiCateg))b3 -0.227 0.107     0.113 -2.01 0.044
const(as.factor(wmiCateg))b4 -0.243 0.095     0.101 -2.39 0.017
const(as.factor(ageCateg))1   0.624 0.121     0.128  4.86 0.000
const(as.factor(ageCateg))2   0.977 0.114     0.125  7.82 0.000
const(as.factor(ageCateg))3   1.500 0.109     0.125 12.00 0.000
\end{verbatim}}
The ability to test the significance of the parameters under the NPH setting (``Supremum-test of significance") and then test if the effects are non-proportional (``Kolmogorov-Smirnov" and "Cramer von Mises" tests), is a powerful tool and is the advantage of reporting of the cumulative estimates as opposed to the traditional estimates for the model.  The {\tt timereg} package also offers the ability to test residuals and other aspects of the NPH model, and to check for goodness-of-fit and parameter significance, all of which are powerful features for model building and analysis.\footnote{For more details on the output above, see \cite{timereg}.}\footnote{While both the {\tt MRH} and {\tt DPpackage} packages offer a \makefxn{summary} function that can be used on the \emrh{} and \lddp{} fitted models, the testing abilities are not as extensive or developed as those for the \timecox{} function in the {\tt timereg} package.}\footnote{We note that the \makefxn{summary} function performs inference on whether the covariate has a non-proportional effect (performed with the Kolmogorov-Smirnov test and also with the Cramer von Mises test.  The Kolmogorov-Smirnov test does not reject the null (that the effect is constant), which disagrees with the Cramer von Mises test as well as the test performed using \makefxn{cox.ph}.  However, the testing of the NPH parameters is outside the scope of this manuscript and we do not explore it further.}

\subsubsection{Graphics}

\begin{SCfigure}
	\centering
	\includegraphics[width=4in]{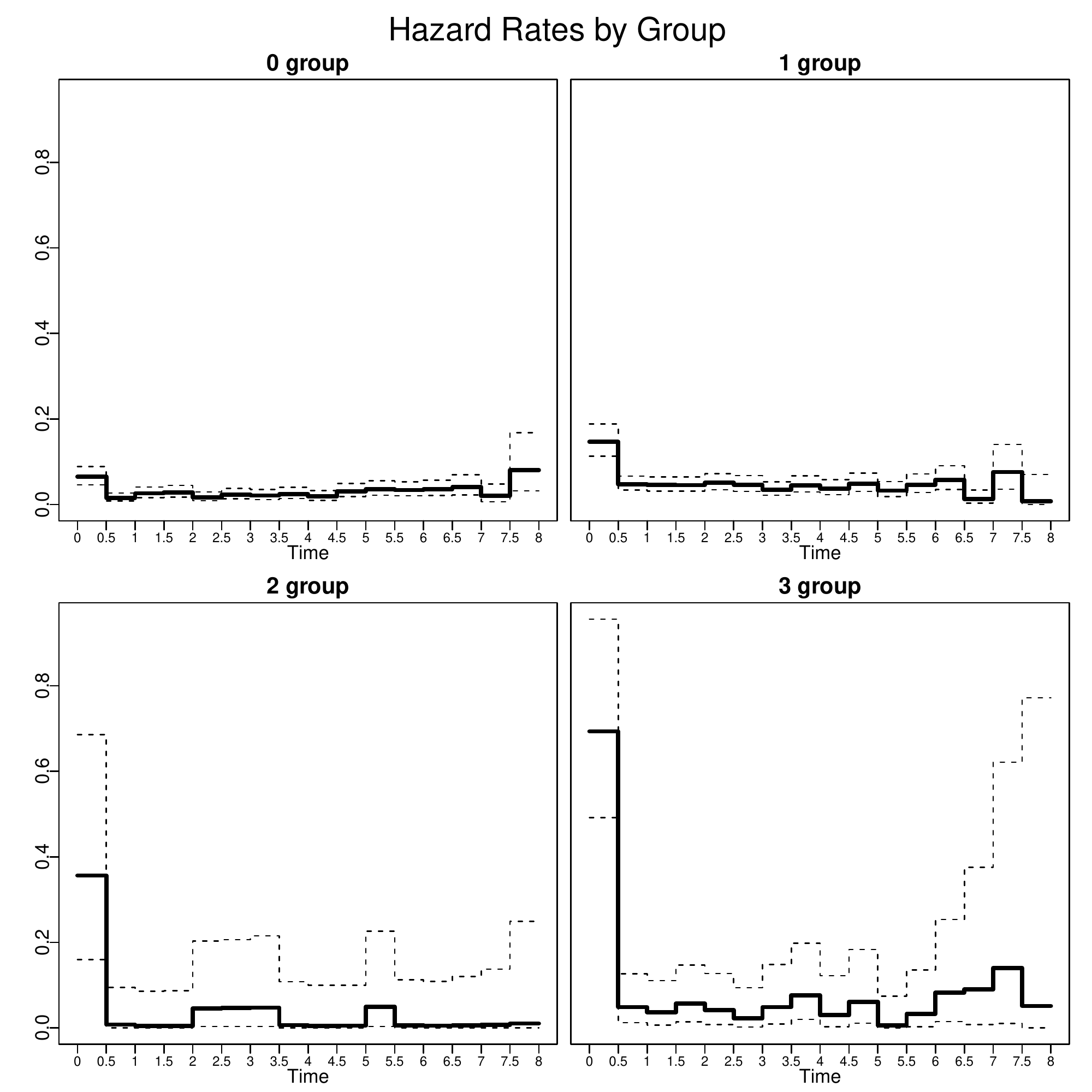}
	\caption{\footnotesize  Plots produced using \makefxn{plot.MRH} on the \emrh{} fitted object, which by default puts all four hazard rates and credible intervals onto one graph. However, they can be separate by setting the ``combine.graphs" option equal to FALSE, which produces the four graphs shown here, and can be easier to read in the instance of multiple NPH covariates.  These graphs display the hazard rate for each of the four non-proportional hazards: no VF or CHF (group 0), CHF only (group 1), VF only (group 2), and CHF + VF (group 3).  Solid lines denote the median of the posterior distribution for the hazard rate, and dashed lines represent the $\alpha$-level (in this case 95\%) credible interval bands.  The graphs show obvious differences in both the magnitude and shape of the hazard rate for each group (note that the y-axis is the same for all four graphs), with the CHF+VF group having the largest hazard rate in the initial 6 months of the study.  Both the VF only and CHF+VF hazard rate estimates have the largest 95\% credible interval bands as they have the smallest number of subjects in each group (2\% and 5\%, respectively).}
\label{fig:default_MRH1}
\end{SCfigure}

\begin{SCfigure}
	\centering
	\includegraphics[width=4in]{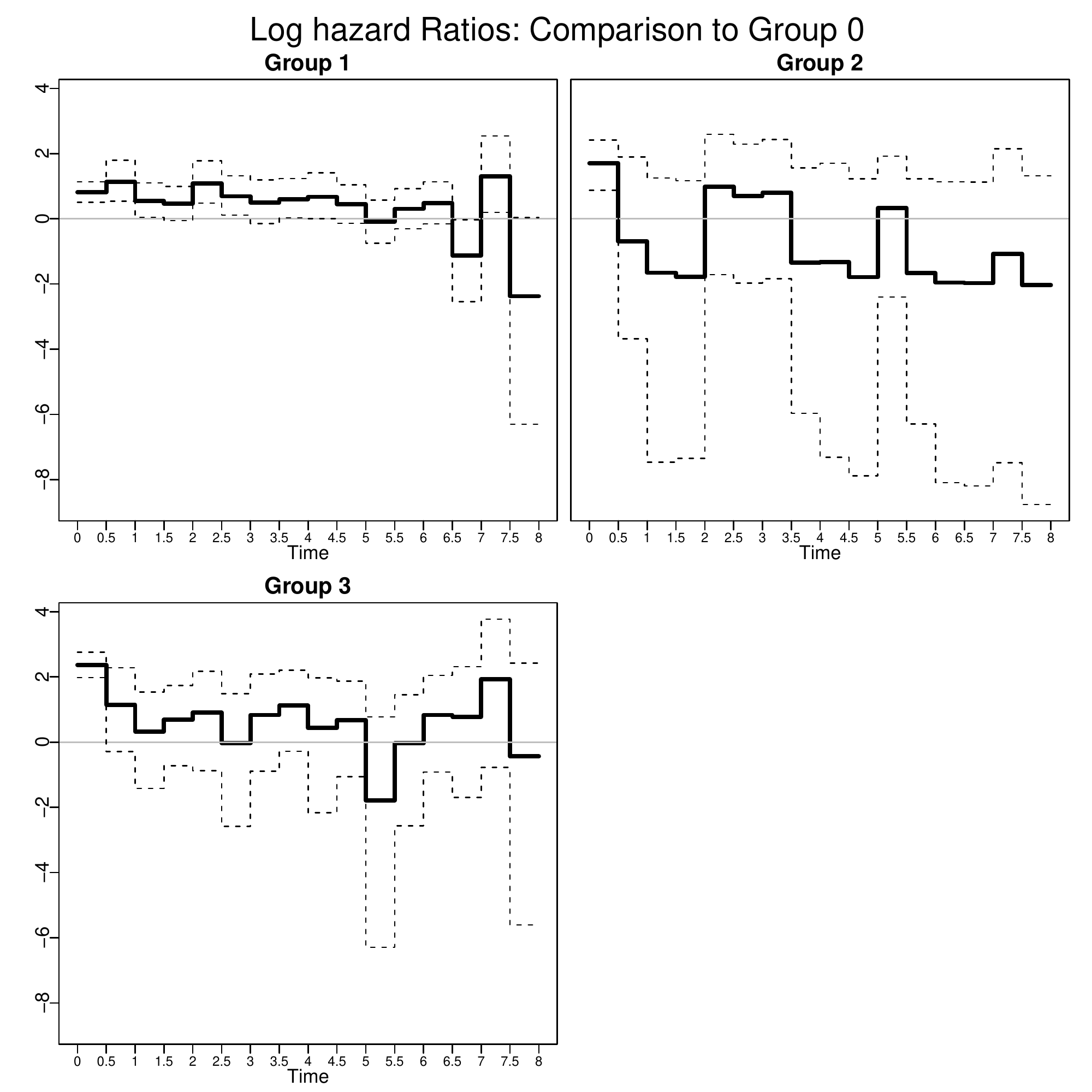}
	\caption{\footnotesize  Plots produced using \makefxn{plot.MRH} on the \emrh{} fitted object, which by default puts all four hazard rates and credible intervals onto one graph.  However, if the ``combine.graphs" option is set to FALSE, and ``plot.type" is set to ``r", the log-ratios (compared to the baseline group) of the NPH covariates are shown on separate figures, as is seen here.  Solid lines denote the median of the posterior distribution for the log-hazard ratio, and dashed lines represent the $\alpha$-level (in this case 95\%) credible interval bands.  The log-hazard ratio estimates show that there is a steady but decreasing effect of CHF only, and, except in the last year of the study, the presence of CHF increased the risk of death after myocardial infarction.  A similar pattern is shown for the CHF+VF group.  Subjects with VF only tended to have a lower risk of survival after myocardial infarction, but the 95\% credible interval bounds are wide due to the small sample size in this group.}
\label{fig:default_MRH2}
\end{SCfigure}

In Figures \ref{fig:default_MRH1} and \ref{fig:default_MRH2}, the default plots produced by using \makefxn{plot.MRH} on the \emrh{} fitted object are shown, with specifications made so that both the hazard rates and hazard ratios are plotted, and also so that the estimates are displayed on separate graphs.\footnote{Code for doing this is as follows:\\
{\small{\tt \# Plot the four hazard rates on separate graphs\\
plot(fit.MRH, combine.graphs = FALSE)\\
\# Plot the hazard ratios on separate graphs\\
plot(fit.MRH, combine.graphs = FALSE, plot.type = `r')}}}
The {\tt MRH} package also allows the user to combine the functions on to one graph, plot the survival or cumulative hazard curves, and change the $\alpha$-level of or omit the credible interval bounds.  

\begin{figure}
	\centering
	\includegraphics[width=6.6in]{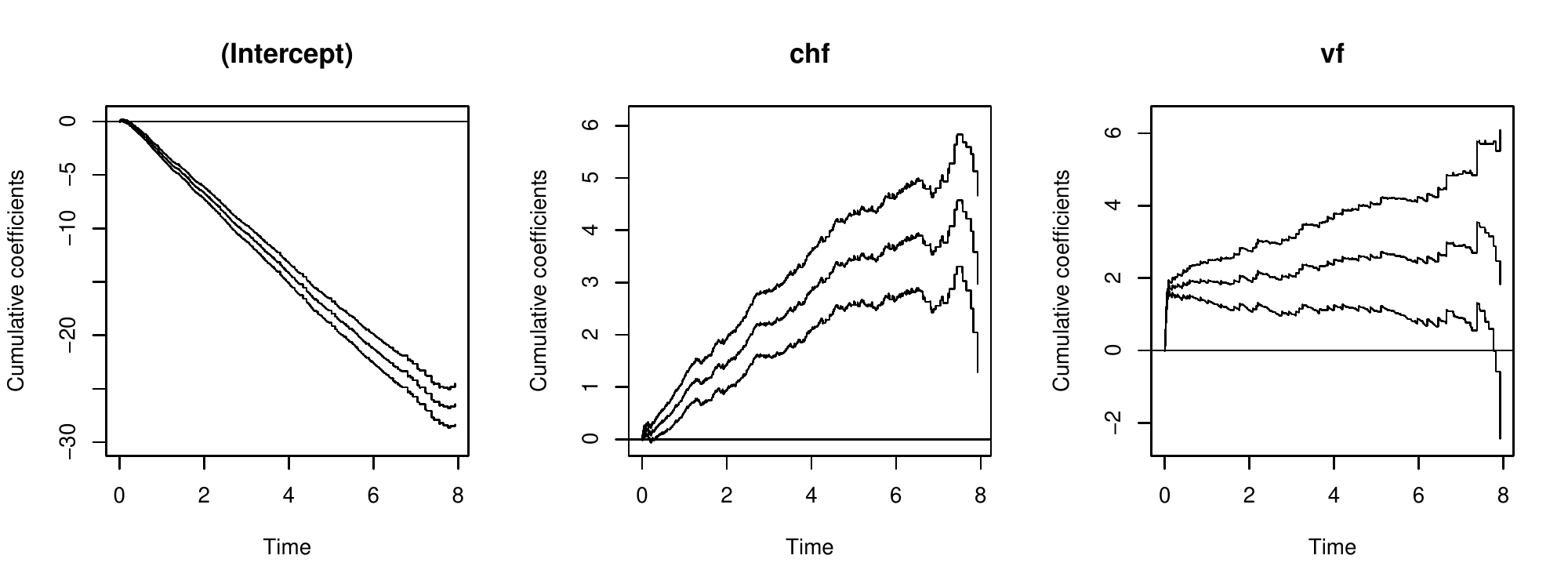}
	\caption{\footnotesize  Default plots produced using \makefxn{plot.timecox} on the \timecox{} fitted survival object.  All estimates shown are cumulative over time.  The left-most figure, labeled ``(Intercept)", is the cumulative estimate for $\alpha_0(t)$ (the log of the baseline hazard rate), the middle figure is the cumulative estimate for CHF, $B_{CFH}(t)$, and the right-most figure is the cumulative estimate for VF, $B_{VF}(t)$.  The middle lines represent the parameter estimate, and the outer lines represent the 95\% confidence intervals.  It is difficult to interpret these parameters in the context of the original problem as the estimates are cumulative.  However, certain strong features are easily recognized.  For example, the estimate of the log-ratio for CHF shows a strong drop in the end, indicating that the impact of this covariate diminishes sharply towards the end of the study.  A similar pattern is observed for the VF variable, although it's impact is more steady over the course of the study.}
\label{fig:default_timecox}
\end{figure}

For contrast, the default \timecox{} plots (created using \makefxn{plot.timecox}\footnote{While the \timecox{} plots appear to accurately display the cumulative parameter estimates, the following internal error message appears when \makefxn{plot.timecox} is used for this particular example:\\
{\tt Error in V[, v] : subscript out of bounds}}) are shown in Figure \ref{fig:default_timecox}. The \makefxn{plot.timecox} function allows the user to change the types of interval bounds shown, combine graphs, and plot only specific variables, if desired.  As noted above, the fitted model parameter estimates look quite different due to the formulation of each model.  

The {\tt bayesSurv} package does not include a plotting function for the fitted \baysurv{} survival object, so plots for this function were not examined.


\section{Discussion}

In this manuscript we have contrasted the implementation of different methodological and computational approaches for the estimation of the hazard rate and the effects of covariates under the proportional and non-proportional settings for right-censored survival data.  Different methodological approaches to survival analysis will carry different advantages and disadvantages in different circumstances, and so a majority of the focus of our survey  was on the features and issues  encountered due to the implementation of the packages.  

One simple addition that greatly increases the user-friendliness of any package is the adoption of some of the standard \R{} input and output formatting.  Users familiar with R will find it easier to fit models, plot results, and print summaries if all packages include standard \R{} \makefxn{formula} for input and and the \makefxn{plot} and \makefxn{summary} functions for output.  In addition, \R{} users encompass a wide spectrum of researchers, ranging from first year students, to medical researchers, to statistical faculty members.  For this reason, documentation, examples, and interpretation for users who are not familiar with the theory behind the model are valued and perhaps necessary for a package to be widely used.

Based on the results of our simulations and analysis of real data sets, the  \bshaz{} method and software seems to be the best in the instance of simple survival models with ``enough" observed failures.  Implementation is easy, as it follows the same input and output patterns as the standard survival model functions (such as \makefxn{survreg} or \makefxn{coxph}), and it reliably estimates the hazard rate as well as the effects of covariates.  However, in more complex situations, the ``best" package is slightly harder to determine.  To estimate a survival model with non-proportional hazards, both the \texttt{MRH} and \texttt{timereg} packages are very user-friendly and are relatively easy to figure out.  The advantage of \emrh{} is that it more reliably estimates the hazard rate, and the estimated covariate effects are reported in a more traditional fashion.  However, the significance of the \timecox{} parameters can be  easily tested using classical approaches, and the time required to fit the \timecox{} model is shorter as it does not require MCMC sampling.  The \lddp{} model, while reliable, is trickier to use due to lack of documentation and idiosyncrasies in the function design, and may require smaller models with few predictors to avoid crashing.  

We also note that users may have a particular preference for certain packages over others depending on whether the underlying methodology is frequentist or Bayesian.  The frequentist models are generally well-known, quick to implement, and perform well with larger data sets.  The Bayesian models allow for more sharing of information and generally have increased power; however, they typically require more computing resources, the user must understand basic (and sometimes more advanced) MCMC sampling techniques, and knowledge of the impact of the priors on the underlying model may be necessary.

While outside the scope of this paper, a number of the packages we examined are also designed to accommodate more complex survival modeling for data that is left-censored, clustered, contains time-varying covariates, and other advanced modeling techniques.  Of the packages we reviewed,\footnote{We also attempted to include the \texttt{dynsurv} package \cite{dynsurvR}, but were unable to get reasonable results, and so did not include it in this manuscript.} \texttt{bayesSurv}, \texttt{BGPhazard}, \texttt{DPpackage}, and \texttt{timereg} all offer very specialized functions that would be challenging to produce from scratch and are difficult (if not impossible) to find elsewhere.  There are many other useful survival packages that we do not discuss for a variety of reasons, such as no estimate of the hazard rate is provided (e.g. \texttt{aftgee} \cite{aftgeeR}), only parametric estimation of the hazard rate is provided (e.g. \texttt{eha} \cite{ehaR}, \texttt{rms} \cite{rmsR}, \texttt{flexsurv} \cite{flexsurvR}), the package is designed for a very specific type of survival analysis (e.g. \texttt{BayHaz} \cite{bayhazR}, \texttt{BaSTA} \cite{bastaR}), or the package is for the analysis of more advanced survival data (e.g. \texttt{JMbayes} \cite{jmbayesR}). 

\section*{Acknowledgements}
The authors would like to acknowledge National Institute of Health grants R21 DA027624-01 and U01 GM087729-03, and the National Science Foundation grants NSF-DEB 1316334 and NSF-GEO 1211668 for partial support. The authors also express gratitude to Steve Scott for very helpful advice with \R{} package building.
The project utilized the Janus supercomputer, which is supported by the National Science Foundation (award number CNS-0821794) and the University of Colorado - Boulder. The Janus supercomputer is a joint effort of the University of Colorado - Boulder, the University of Colorado - Denver, and the National Center for Atmospheric Research. Janus is operated by the University of Colorado - Boulder.

\bibliographystyle{plain}

\begin{thebibliography}{10}

\bibitem{AofNA}
O.~Aalen.
\newblock Nonparameteric inference for a family of counting processes.
\newblock {\em {Annals of Statistics}}, 6:701--726, 1978.

\bibitem{Bouman2}
P.~Bouman, J.~Dignam, V.~Dukic, and X.L. Meng.
\newblock A multiresolution hazard model for multi-center survival studies:
  Application to {T}amoxifen treatment in early stage breast cancer.
\newblock {\em {Journal of the American Statistical Association}},
  102:1145--1157, 2007.

\bibitem{Bouman}
P.~Bouman, V.~Dukic, and X.L. Meng.
\newblock Bayesian multiresolution hazard model with application to an {AIDS}
  reporting delay study.
\newblock {\em {Statistica Sinica}}, 15:325--357, 2005.

\bibitem{ehaR}
G.~Brostr\''om.
\newblock eha: Event history analysis.
\newblock \url{http://cran.r-project.org/web/packages/eha/index.html}.

\bibitem{presmooth1}
R.~Cao and M.A. Jac\'ome.
\newblock Presmoothed kernel density estimation for censored data.
\newblock {\em Journal of Nonparametric Statistics}, 16:289--309, 2004.

\bibitem{presmooth3}
R.~Cao and I.~L\'opez{-}de{-}Ullibarri.
\newblock Product-type and presmoothed hazard rate estimators with censored
  data.
\newblock {\em Test}, 16:355--382, 2007.

\bibitem{presmooth2}
R.~Cao, I.~L\'opez{-}de{-}Ullibarri, P.~Janssen, and N.~Varaverbeke.
\newblock Presmoothed {K}aplan-{M}eier and {N}elson-{A}alen estimators.
\newblock {\em Journal of Nonparametric Statistics}, 17:31--56, 2005.

\bibitem{Yprune}
Y.~Chen, Y.~Hagar, J.~Dignam, and V.~Dukic.
\newblock Pruned {M}ultiresolution {H}azard ({PMRH}) models for time-to-event
  data.
\newblock {\em Bayesian Analysis}, In Review.

\bibitem{aftgeeR}
S.~Chiou, S.~Kang, and J.~Yan.
\newblock Package `aftgee'.
\newblock \url{http://cran.r-project.org/web/packages/aftgee/index.html}.

\bibitem{bastaR}
F.~Colchero, O.~Jones, and M.~Rebke.
\newblock Package `{B}a{STA}'.
\newblock \url{http://cran.r-project.org/web/packages/BaSTA/index.html}.

\bibitem{CoxPH}
D.R. Cox.
\newblock Regression models and life-tables.
\newblock {\em {Journal of the Royal Statistical Society - Series B}},
  34:187--220, 1972.

\bibitem{Iorio}
M.~{De Iorio}, W.O. Johnson, P.~M{\"u}ller, and G.L. Rosner.
\newblock Bayesian nonparametric nonproportional hazards survival modeling.
\newblock {\em {Biometrics}}, 65:762--771, 2009.

\bibitem{IorioAnova}
M.~{De Iorio}, P.~M{\"u}ller, G.L. Rosner, and S.N. MacEachern.
\newblock An {ANOVA} model for dependent random measures.
\newblock {\em {Journal of the American Statistical Association}}, 99:202--215,
  2004.

\bibitem{Dignam}
J.~Dignam, V.~Dukic, S.~Anderson, E.~Mamounas, D.~Wickerham, and N.~Wolmark.
\newblock Hazard of recurrence and adjuvant treatment effects over time in
  lymph node-negative breast cancer.
\newblock {\em Breast Cancer Research and Treatment}, 116:595--602, 2009.

\bibitem{Dukic}
V.~Dukic and J.~Dignam.
\newblock Bayesian hierarchical multiresolution hazard model for the study of
  time-dependent failure patterns in early stage breast cancer.
\newblock {\em Bayesian Analysis}, 2:591--610, 2007.

\bibitem{BGPhazardR}
J.A. Garcia and L.E. Nieto-Barajas.
\newblock {BGP}hazard package in {R}.
\newblock \url{http://cran.r-project.org/web/packages/BGPhazard/index.html}.

\bibitem{muhaz2}
O.~Gefeller and H.~Dette.
\newblock Nearest neighbor kernel estimation of the hazard function from
  censored data.
\newblock {\em Journal of Statistical Computation and Simulation}, 43:93--101,
  1992.

\bibitem{BayesianMixedMods}
A.~Gelman, J.B. Carlin, H.S. Stern, and D.B. Rubin.
\newblock {\em Bayesian Data Analysis, 2nd edition}.
\newblock Chapman and Hall/CRC, Boca Raton, 2004.

\bibitem{gelmanrubin}
A.~Gelman and D.B. Rubin.
\newblock Inference from iterative simulation using multiple sequences.
\newblock {\em Statistical Science}, 7:457--511, 1992.

\bibitem{presmooth4}
W.~Gonzalez-Manteiga, R.~Cao, and J.S. Marron.
\newblock Bootstrap selection of the smoothing parameter in nonparametric
  hazard rate estimation.
\newblock {\em {Journal of the American Statistical Association}},
  91:1130--1140, 1996.

\bibitem{MRHR}
Y.~Hagar, Y.~Chen, and V.~Dukic.
\newblock {MRH} package in {R}.
\newblock \url{http://cran.r-project.org/web/packages/MRH/index.html}.

\bibitem{rmsR}
F.~{Harrell Jr.}
\newblock Package `rms'.
\newblock \url{http://cran.r-project.org/web/packages/rms/index.html}.

\bibitem{muhazR}
K.~Hess and R.~Gentleman.
\newblock muhaz package in {R}.
\newblock \url{http://cran.r-project.org/web/packages/muhaz/index.html}.

\bibitem{muhaz3}
K.R. Hess and D.M. Serachitopol.
\newblock Hazard function estimators: a simulation study.
\newblock {\em Statistics in Medicine}, 18:3075--3088, 1999.

\bibitem{flexsurvR}
C.~Jackson.
\newblock Package `flexsurv'.
\newblock \url{http://cran.r-project.org/web/packages/flexsurv/index.html}.

\bibitem{presmooth5}
M.A. Jac\'ome, I.~Gijbels, and R.~Cao.
\newblock Comparison of presmoothing methods in kernel density estimation under
  censoring.
\newblock {\em Computational Statistics}, 23:381--406, 2008.

\bibitem{dppkgR}
A.~Jara, T.~Hanson, F.~Quintana, P.~M\''ueller, and G.~Rosner.
\newblock Package `dppackage'.
\newblock \url{http://cran.r-project.org/web/packages/DPpackage/index.html}.

\bibitem{DPpkgJSS}
A.~Jara, T.~Hanson, F.~Quintana, P.~M\"{u}eller, and G.~Rosner.
\newblock {DP}package: {B}ayesian semi- and nonparametric modeling in {R}.
\newblock {\em Journal of Statistical Software}, 40, 2011.

\bibitem{tracedataset}
G.V. Jensen, C.~Torp-Pedersen, P.~Hildebrandt, L.~Kober, F.~E. Nielsen,
  T.~Melchior, T.~Joen, and P.K. Andersen.
\newblock Does in-hospital ventricular fibrillation affect prognosis after
  myocardial infarction?
\newblock {\em European Heart Journal}, 18:919--924, 1997.

\bibitem{KMest}
E.~Kaplan and P.~Meier.
\newblock Nonparametric estimation from incomplete observations.
\newblock {\em {Journal of the American Statistical Association}}, 53:457--481,
  1958.

\bibitem{bayessurvR}
A.~Kom\'arek.
\newblock Package `bayessurv'.
\newblock \url{http://cran.r-project.org/web/packages/bayesSurv/index.html}.

\bibitem{bayessurvKomarekthesis}
A.~Kom\'arek.
\newblock {\em Accelerated Failure Time Models for Multivariate
  Interval-Censored Data with Flexible Distributional Assumptions}.
\newblock PhD thesis, Katholieke Universiteit Leuven, Faculteit Wetenschappen,
  2006.

\bibitem{bayessurvKomarek}
A.~Kom\'arek and E.~Lesaffre.
\newblock Bayesian accelerated failure time model for correlated
  interval-censored data with a normal mixture as an error distribution.
\newblock {\em Statistica Sinica}, 17:549--569, 2007.

\bibitem{bayhazR}
L.~{La Rocca}.
\newblock Package `{B}ay{H}az'.
\newblock \url{http://cran.r-project.org/web/packages/BayHaz/index.html}.

\bibitem{bshazard1}
Y.~Lee, J.A. Nelder, and Y.~Pawitan.
\newblock {\em Generalized Linear models with random effects: {U}nified
  analysis via {H}-likelihood}.
\newblock Chapman \& Hall/CRC, Boca Raton, FL, 2006.

\bibitem{createRpkg}
F.~Leisch.
\newblock Creating {R} packages: {A} tutorial.
\newblock
  \url{http://cran.r-project.org/doc/contrib/Leisch-CreatingPackages.pdf}.

\bibitem{survpresmoothR}
I~L\'opez{-}de{-}Ullibarri and M.~Jac\'ome.
\newblock surv{P}resmooth package in {R}.
\newblock
  \url{http://cran.r-project.org/web/packages/survPresmooth/index.html}.

\bibitem{presmoothJSS}
I.~L\'opez{-}de{-}Ullibarri and M.~Jac\'ome.
\newblock survpresmooth: {A}n {R} package for pre smoothed estimation in
  survival analysis.
\newblock {\em Journal of Statistical Software}, 43, 2013.

\bibitem{cancerdataset}
C.L Loprinzi, J.A. Laurie, H.S. Wieand, J.E. Krook, P.J. Novotny, J.W. Kugler,
  J.~Bartel, M.~Law, M.~Bateman, and N.E. Klatt.
\newblock Prospective evaluation of prognostic variables from patient-completed
  questionnaires. {N}orth {C}entral {C}ancer {T}reatment {G}roup.
\newblock {\em Journal of Clinical Oncology}, 12:601--607, 1994.

\bibitem{timereg}
T.~Martinussen and T.H. Scheike.
\newblock {\em Dynamic Regression Models for Survival Data}.
\newblock Springer, New York, NY, 2006.

\bibitem{muhaz1}
H.G. Mueller and J.L. Wang.
\newblock Hazard rate estimation under random censoring with varying kernels
  and bandwidths.
\newblock {\em Biometrics}, 50:61--76, 1994.

\bibitem{nadaraya}
E.A. Nadaraya.
\newblock On estimating regression.
\newblock {\em Theory of Probability and its Applications}, 9:141--142, 1964.

\bibitem{NofNA}
W.~Nelson.
\newblock Theory and applications of hazard plotting for censored failure data.
\newblock {\em TechnometricsS}, 15:945--966, 1972.

\bibitem{nieto}
L.~Nieto-Barajas and S.~Walker.
\newblock Markov beta and gamma processes for modeling hazard rates.
\newblock {\em Scandinavian Journal of Statistics}, 29:413--424, 2002.

\bibitem{BGPhazard}
L.E. Nieto-Barajas.
\newblock Discrete time markov gamma processes for modeling hazard rates in
  survival analysis.
\newblock {\em Bulletin of the International Statistical Institute 54th
  Session. Berlin.}, CD-ROM, 2003.

\bibitem{ecog}
M.~Oken, R.~Creech, D.~Tormey, J.~Horton, T.~Davis, E.~McFadden, and
  P.~Carbone.
\newblock Toxicity and response criteria of the {E}astern {C}ooperative
  {O}ncology {G}roup.
\newblock {\em American Journal of Clinical Oncology}, 5:649--656, 1982.

\bibitem{bshazard2}
Y.~Pawitan.
\newblock {\em In all likelihood: statistical modelling and inference using
  likelihood}.
\newblock Oxford University Press, United Kingdom, 2013.

\bibitem{Rref}
{R Development Core Team}.
\newblock {\em R: A Language and Environment for Statistical Computing}.
\newblock R Foundation for Statistical Computing, Vienna, Austria, 2008.
\newblock {ISBN} 3-900051-07-0.

\bibitem{bshazardR}
P.~Rebora, A.~Salim, and M.~Reilly.
\newblock bshazard package in {R}.
\newblock \url{http://cran.r-project.org/web/packages/bshazard/index.html}.

\bibitem{bshazardMan}
P.~Rebora, A.~Salim, and M.~Reilly.
\newblock bshazard: {A} flexible tool for nonparametric smoothing of the hazard
  function.
\newblock {\em The R Journal}, 6, 2014.

\bibitem{bayessurvRichardson}
S.~Richardson and P.J. Green.
\newblock On bayesian analysis of mixtures with unknown number of components
  (with discussion).
\newblock {\em JRSSB}, 59:731--792, 1997.

\bibitem{jmbayesR}
D.~Rizopoulos.
\newblock Package `{JM}bayes'.
\newblock \url{http://cran.r-project.org/web/packages/JMbayes/index.html}.

\bibitem{timeregPkg}
T.~Scheike.
\newblock timereg package in {R}.
\newblock \url{http://cran.r-project.org/web/packages/timereg/timereg.pdf}.

\bibitem{ScheikeCR}
T.~Scheike and M.~Zhang.
\newblock Analyzing competing risk data using the {R} timereg package.
\newblock {\em Journal of Statistical Software}, 38, 2011.

\bibitem{ypmodelR}
J.~Sun and S.~Yang.
\newblock {YP}model package in {R}.
\newblock \url{http://cran.r-project.org/web/packages/YPmodel/index.html}.

\bibitem{tannerwong}
M.~Tanner and W.~Wong.
\newblock The estimation of the hazard function from randomly censored data by
  the kernel method.
\newblock {\em The Annals of Statistics}, 11:989--993, 1983.

\bibitem{coxphsw}
T.M. Therneau.
\newblock Fit {C}ox proportional hazards regression model in {R}.
\newblock
  \url{http://stat.ethz.ch/R-manual/R-patched/library/survival/html/coxph.html%
}.

\bibitem{survpkg}
T.M. Therneau.
\newblock Survival analysis.
\newblock \url{http://cran.r-project.org/web/packages/survival/survival.pdf}.

\bibitem{dynsurvR}
X.~Wang, J.~Yan, and M.~Chen.
\newblock dynsurv package in {R}.
\newblock \url{http://cran.r-project.org/web/packages/dynsurv/index.html}.

\bibitem{watson}
G.S. Watson.
\newblock Smooth regression analysis.
\newblock {\em Sankyh{\u a}: The Indian Journal of Statistics, Series A},
  26:359--372, 1964.

\bibitem{ypmodel1}
S.~Yang and R.L. Prentice.
\newblock Improved logrank-tests for survival data using adaptive weights.
\newblock {\em {Biometrics}}, 66:30--38, 2010.

\bibitem{ypmodel2}
S.~Yang and Y.~Zhao.
\newblock Checking the short-term and long-term hazard ratio model for survival
  data.
\newblock {\em Scandinavian Journal of Statistics}, 39:554--567, 2012.

\end{thebibliography}

\end{document}